\documentclass[english]{article}
\usepackage[utf8]{inputenc}
\usepackage{amsmath}
\usepackage{amssymb}
\usepackage{mathrsfs}
\usepackage{cite}
\usepackage{multirow}
\usepackage{framed}
\usepackage[colorlinks=true,urlcolor=blue,anchorcolor=blue,citecolor=blue,filecolor=blue,linkcolor=blue,menucolor=blue,linktocpage=true,pdfproducer=medialab,pdfa=true]{hyperref}
\usepackage{calc}
\usepackage{breqn}
\usepackage{color}
\usepackage{float}

\definecolor{camouflagegreen}{rgb}{0.47, 0.53, 0.42}

\DeclareMathOperator{\Li}{Li}

\allowdisplaybreaks

\makeatletter

\@addtoreset{equation}{section}

\def\draftdate{\relax}
\def\mda{\relax}
\def\mua{\relax}
\def\mla{\relax}
\def\draft{
\def\thtystars{******************************}
\def\sixtystars{\thtystars\thtystars}
\typeout{}
\typeout{\sixtystars**}
\typeout{* Draft mode!
         For final version remove \protect\draft\space in source file *}
\typeout{\sixtystars**}
\typeout{}
\def\draftdate{\today}
\def\mua{\marginpar[\boldmath\hfil$\uparrow$]%
                   {\boldmath$\uparrow$\hfil}%
                    \typeout{marginpar: $\uparrow$}\ignorespaces}
\def\mda{\marginpar[\boldmath\hfil$\downarrow$]%
                   {\boldmath$\downarrow$\hfil}%
                    \typeout{marginpar: $\downarrow$}\ignorespaces}
\def\mla{\marginpar[\boldmath\hfil$\rightarrow$]%
                   {\boldmath$\leftarrow $\hfil}%
                    \typeout{marginpar: $\leftrightarrow$}\ignorespaces}
\overfullrule 5pt
\oddsidemargin -12mm
\marginparwidth 29mm
}

\def\stars{\strut\leaders\hbox{*}\hfill\strut}
\def\starline{\hfil\strut\hfil\hbox to \textwidth {\stars}\hfil}

\newcommand\refr[1]      {ref.\,\cite{#1}}
\newcommand\refrs[1]    {refs.\,\cite{#1}}

\newcommand\eqn[1]     {eq.\,(\ref{#1})}

\newcommand\eqnss[2]   {eqs.\,(\ref{#1})--(\ref{#2})}

\newcommand\fig[1]     {fig.\,{\ref{#1}}}
\newcommand\figs[2]     {figs.\,{\ref{#1}} and~\ref{#2}}

\newcommand\sect[1]    {sec.\,{\ref{#1}}}

\newcommand\appx[1]     {appendix~\ref{#1}}

\newcommand\tab[1]     {table~\ref{#1}}
\newcommand\tabs[2]     {tables~\ref{#1} and~\ref{#2}}

\newcommand\nn         {\nonumber}

\def\beq{\begin{equation}}
\def\eeq{\end{equation}}
\def\bsp#1\esp{\begin{split}#1\end{split}}
\def\bal#1\eal{\begin{align}#1\end{align}}

\def\beeq{\begin{eqnarray}}
\def\eeeq{\end{eqnarray}}
\newcommand\bom[1]     {{\mbox{\boldmath $#1$}}}

\newcommand\as         {\ensuremath{\alpha_{\mathrm{s}}}}

\newcommand{\CA}       {C_{\mathrm{A}}}

\newcommand{\Nf}       {n_{\mathrm{f}}}

\newcommand\msbar      {\ensuremath{{\overline {\rm MS}}}}

\newcommand\Oe[1]      {\ensuremath{\mathrm O(\eps^{#1})}}

\newcommand{\ep}       {\varepsilon}
\newcommand{\eps}      {\varepsilon}

\newcommand{\rd}       {{\mathrm{d}}}
\newcommand{\PS}[1]    {\rd\phi_{#1}}
\newcommand\tsig[1]    {\sigma^{\mathrm{#1}}}
\newcommand\tsigk[4]    {\sigma^{{\rm #1}^{#2} {\rm #3}^{#4}}}
\newcommand\dsig[1]    {\rd\sigma^{{\rm #1}}}
\newcommand\dsigk[4]    {\rd\sigma^{{\rm #1}_{#2} {\rm #3}_{#4}}}

\newcommand\dsiga[2]   {\rd\sigma^{{\rm #1,A}_{\scriptscriptstyle #2}}}

\newcommand\la         {\langle}
\newcommand\ra         {\rangle}

\newcommand{\cA}       {{\cal A}}
\newcommand{\cII}[1] {{\cal I}\kern-4pt *\kern-4pt{\cal I}_{#1}}
\newcommand{\cIJ}     {{\cal I}\kern-4pt *\kern-4pt{\cal J}}
\newcommand{\cJJ}[1] {{\cal J}\kern-4pt *\kern-4pt{\cal J}_{#1}}
\newcommand{\cJI}     {{\cal J}\kern-4pt *\kern-4pt{\cal I}}
\newcommand{\cJK}     {{\cal J}\kern-4pt *\kern-4pt{\cal K}}
\newcommand{\cKJ}[1] {{\cal K}\kern-4pt *\kern-4pt{\cal J}_{#1}}
\newcommand{\cKI}     {{\cal K}\kern-4pt *\kern-4pt{\cal I}}
\newcommand{\cKK}     {{\cal K}\kern-4pt *\kern-4pt{\cal K}}

\newcommand\SME[3]     {|{\cal M}_{#1}^{#2}{#3}|^2}

\newcommand\braket[6]     {\la {\cal M}_{#1}^{#2}#3|{\cal M}_{#4}^{#5}#6\ra}

\newcommand{\mom}[1]   {\{p\}^{#1}}

\newcommand{\momt}[1]   {\{\ti{p}\}^{#1}}
\newcommand{\momh}[1]   {\{\ha{p}\}^{#1}}

\newcommand{\bSCS}[1]  {\bom{\mathrm C}\kern-2pt\bom{\mathrm S}_{#1}}

\newcommand{\cC}[2]    {{\cal C}_{#1}^{#2}}
\newcommand{\cS}[2]    {{\cal S}_{#1}^{#2}}

\newcommand{\cSCS}[2]  {{\cal C}\kern-2pt{\cal S}_{#1}^{#2}}

\newcommand{\IcSCS}[2]  {\mathrm{C}\kern-2pt\mathrm{S}_{#1}^{#2}}
\newcommand{\bI}       {\bom{I}}

\newcommand{\ti}[1]    {\tilde{#1}}

\newcommand{\ha}[1]    {\hat{#1}}

\def\s12{s_{12}}

\begin{document}

%%%
%%% Title page
%%%

\begin{titlepage}
%\renewcommand{\thefootnote}{\fnsymbol{footnote}}

% Title
\noindent
DESY-24-211 \hfill December 2024\\
BONN-TH-2024-18
\vspace{0.6cm}
\begin{center}
{\LARGE \bf 
{\tt NNLOCAL}: completely local subtractions for color-singlet production in hadron collisions
}
\vspace{1.0cm}

\large
V. Del Duca$^{\, a}$, C. Duhr$^{\, b}$, L. Fek\'esh\'azy$^{\, c,d}$, F. Guadagni$^{\, e}$, P. Mukherjee$^{\, c}$,\\[0.25em] G. Somogyi$^{\, f}$, F. Tramontano$^{\, g}$ and S.~Van Thurenhout$^{\, f}$\\
\vspace{0,5cm}
\normalsize
{\it $^{\, a}$INFN, Laboratori Nazionali di Frascati, 00044 Frascati (RM), Italy}\\
{\it $^{\, b}$Bethe Center for Theoretical Physics, Universit\"at Bonn, D-53115, Germany}\\
{\it $^{\, c}$II. Institut f\"ur Theoretische Physik, Universit\"at Hamburg, Luruper Chaussee
149, 22761, Hamburg, Germany}\\
{\it $^{\, d}$Institute for Theoretical Physics, ELTE E\"otv\"os Lor\'and University, P\'azm\'any P\'eter s\'et\'any
1/A, 1117, Budapest, Hungary}\\
{\it $^{\, e}$Physik-Institut, Universit\"at Z\"urich, 8057 Z\"urich, Switzerland}\\
{\it $^{\, f}$HUN-REN Wigner Research Centre for Physics, Konkoly-Thege Mikl\'os u. 29-33, 1121 Budapest, Hungary}\\
{\it $^{\, g}$Dipartimento di Fisica Ettore Pancini, Universit\`a di Napoli Federico II and INFN - Sezione di
Napoli, Complesso Universitario di Monte Sant’Angelo Ed. 6, Via Cintia, 80126 Napoli, Italy}
\vspace{1.4cm}

{\large \bf Abstract}
\vspace{-0.2cm}
\end{center}
We present {\tt NNLOCAL}, a proof-of-concept parton-level Monte Carlo program implementing the extension of the completely local subtraction scheme CoLoRFulNNLO to the case of color-singlet production in hadron collisions. We have built general local subtraction terms that regularize all single and double unresolved infrared singularities in real radiation phase space. The subtractions are then integrated fully analytically to the required order in the parameter of dimensional regularization. Combining the integrated counterterms with the virtual contributions we demonstrate the cancellation of all infrared poles explicitly. We validate our procedure by computing the fully differential cross section for the production of a Higgs boson at the LHC in an effective field theory with gluons only. Our code provides the first public implementation of a completely local analytic subtraction scheme at next-to-next-to-leading order accuracy.
\end{titlepage}
\clearpage

%%%
%%% TOC here
%%%

\tableofcontents

%\renewcommand{\thefootnote}{\fnsymbol{footnote}}

%%%%%%%%%%%%%%%%%%%%%%%%%%%%%%%%%%%%%%%%%%%%%%%%%%%%%%%%%%%%
%%%%%%%%%%%%%%%%%%%%%%%%%%%%%%%%%%%%%%%%%%%%%%%%%%%%%%%%%%%%

%%%
%%% Introduction
%%%

\section{Introduction}
\label{sec:intro}

The Standard Model (SM) of particle physics provides a
well-established description of the known particles and their
interactions over a wide range of energies. While the last major
ingredient, the Higgs boson, was discovered at the LHC over a decade
ago \cite{ATLAS:2012yve,CMS:2012qbp}, certain mysteries within the
model still remain. For example, while we think we understand some of
the more general features of the Higgs field, certain details such as
the exact shape of the potential and the nature of self-interactions
still remain puzzling. To unravel such mysteries, current as well as
planned high-precision experiments are vital. Examples of the latter
include the high-luminosity phase of the LHC \cite{HL-LHC} in the near
future and the possibility of the FCC-hh
\cite{Golling:2016gvc,Contino:2016spe} in the far future. Besides the
more in-depth study of the SM itself, high-precision physics may also
prove crucial for finding signs of physics {\em beyond} the
Standard Model, which is motivated, e.g., by the matter-anti-matter
asymmetry and the non-zero neutrino masses.

At the theoretical level, one important aspect of increasing precision
involves computing higher-order perturbative corrections to physical
observables such as scattering cross sections. This generically
requires one to take into account additional emissions compared to the
Born process. The corresponding radiated partons can either be virtual or real, leading
to additional loops or legs in the Feynman diagram expansion. As is
well known, the resulting diagrams develop singularities in several
regions of phase space. In particular, loops give rise to both ultraviolet
(UV) and infrared (IR) divergences and at the same time the production of 
unresolved real partons leads to IR singularities. While UV divergences are removed once and for all by renormalization, there is
no unique method for treating IR singularities. This is highlighted by the
fact that, although the problem is considered to be solved at
next-to-leading order (NLO) \cite{Frixione:1995ms,Catani:1996vz}, the
computation of next-to-next-to-leading order (NNLO) corrections is a
highly active field of research with several approaches available in
the literature
\cite{Catani:2007vq,Gaunt:2015pea,Gehrmann-DeRidder:2005btv,Czakon:2014oma,Caola:2017dug,Cacciari:2015jma,Magnea:2018hab,Herzog:2018ily}. Conceptually,
a particularly appealing approach lies in the construction of a local
subtraction scheme. Such schemes are deeply connected to the universal
nature of the IR structure of squared matrix elements in quantum chromodynamics (QCD). Specifically, the basic idea is to define {\em approximate
  cross sections} that match the point-wise singular behavior of the
real-emission partonic cross sections. When these approximations are
subtracted, one obtains an expression which is regular as real
emissions become unresolved. Moreover, the subtraction terms need to
be integrated over the phase space of the unresolved emissions and
added back, which in turn takes care of the singularities coming from
virtual contributions. Hence, one ends up with blocks of phase space
integrals which are separately finite and so can be evaluated
numerically.

The construction of the approximate cross sections is directly
inspired by QCD factorization and the universal nature of the IR
structure of QCD matrix elements
\cite{Collins:1985ue,Collins:1989gx,Collins:2004nx}. In particular,
exploiting universality, one can hope to construct the required
approximate cross sections in a process- and observable-independent
fashion. However, while the IR structure of QCD has been extensively
studied at NNLO
\cite{Campbell:1997hg,Catani:1999ss,Bern:1999ry,Catani:2000pi},
carrying out the full program of constructing universal local
approximate cross sections has proven to be surprisingly
challenging. In the present paper we address the extension of the
CoLoRFulNNLO subtraction scheme
\cite{Somogyi:2005xz,Somogyi:2006db,Somogyi:2006da,DelDuca:2015zqa,DelDuca:2016csb,DelDuca:2016ily,Somogyi:2020mmk}
to hadron-hadron collisions. This method was applied previously to
processes with colorless initial states and starts by considering the
known IR limits of QCD amplitudes. These are then promoted to true
subtraction terms by carefully defining the momenta entering the
factorized matrix elements and specifying the various quantities
entering the IR factorization formul\ae. To avoid multiple
subtractions in regions of phase space where limits overlap,
subtraction terms based on iterated limit formul\ae\ are
constructed. The approximate cross sections obtained in this way are
completely local. In particular, the point-wise convergence of the sum
of subtraction terms to the real radiation contribution can be
demonstrated explicitly. Moreover, the subtraction terms can be
integrated analytically over the unresolved emissions, up to the
appropriate order of $\ep$ in dimensional regularization in $d=4-2\ep$
dimensions. Combining the results with the virtual contributions, the
complete cancellation of $\ep$-poles can be shown. We demonstrate the
viability of this approach by presenting a proof-of-concept
parton-level Monte Carlo code for computing NNLO corrections to
color-singlet production in hadron collisions. For now, we focus on
Higgs boson production in the Higgs effective field theory (HEFT)
approximation
\cite{Ellis:1975ap,Voloshin:1985tc,Shifman:1988zk,Dawson:1990zj,Djouadi:1991tka,Graudenz:1992pv,Spira:1995rr}
without light quarks, considering only the fully gluonic
subprocess. We emphasize that this is not a restriction on the
structure of the subtraction scheme. In fact, the fully gluonic
subprocess has a highly non-trivial IR structure and receives
contributions from {\em all} possible types of IR singularities for
color-singlet production. As such, it provides an ideal testing ground
for setting up the subtraction without having to worry about technical
complications which are irrelevant for our current purposes.

The paper is organized as follows. In \sect{sec:local} we review
the general setup of the CoLoRFulNNLO subtraction scheme. Our goal
here is to present the overall picture without entering into the
rather elaborate details of the precise definition of each subtraction
term and its corresponding integration, which will be given
elsewhere. Instead, we provide the results in the form of a publicly
available computer code called {\tt NNLOCAL}. The latter will be
introduced in \sect{sec:nnlocal} and constitutes a
proof-of-concept implementation of the method in a particle-level
Monte Carlo program. Finally, in \sect{sec:conc} we present our conclusions and outlook.

%%%%%%%%%%%%%%%%%%%%%%%%%%%%%%%%%%%%%%%%%%%%%%%%%%%%%%%%%%%%
%%%%%%%%%%%%%%%%%%%%%%%%%%%%%%%%%%%%%%%%%%%%%%%%%%%%%%%%%%%%

%%%
%%% Local subtraction at NNLO
%%%

\section{Local subtraction at NNLO}
\label{sec:local}

% General subtraction procedure

\subsection{General subtraction procedure}
\label{ssec:sub-proc}

We consider the production of a colorless final state $X$ in hadron-hadron collisions, $A(p_A)+B(p_B) \to X(p_X)$. Cross sections for such processes are computed by convoluting parton density functions (PDFs) with partonic cross sections,
\beq
\sigma(p_A,p_B) = \sum_{a,b}
	\int_0^1 \rd x_a f_{a/A}(x_a,\mu_F^2) \int_0^1 \rd x_b f_{b/B}(x_b,\mu_F^2)\, 
    \tsig{}_{ab}(p_a,p_b;\mu_F^2)\,.
\eeq
Here the partonic momenta are $p_a = x_a p_A$ and $p_b = x_b p_B$ and the sum is over parton flavors. The partonic cross section can be computed in perturbation theory and up to NNLO accuracy it reads
\beq
\tsig{}_{ab}(p_a,p_b;\mu_F^2) = 
    \tsig{LO}_{ab}(p_a,p_b;\mu_R^2,\mu_F^2)
    + \tsig{NLO}_{ab}(p_a,p_b;\mu_R^2,\mu_F^2)
    + \tsig{NNLO}_{ab}(p_a,p_b;\mu_R^2,\mu_F^2)
    + \ldots
    \,.
\eeq
In the following, the dependence of the cross sections on partonic momenta, as well as on the renormalization and factorization scales $\mu_R^2$ and $\mu_F^2$, will be suppressed. 

The leading order (LO) contribution is simply the integral of the fully differential Born cross section over the phase space of the produced color-singlet state, 
\beq
\tsig{LO}_{ab} = \int_X \dsig{B}_{ab}\, J_X\,.
\label{eq:tsigLO}
\eeq
Here $J_X$ is the value of some infrared and collinear-safe measurement function $J$ evaluated on the Born phase space. More generally, we will denote by $J_{X+n}$ the value of $J$ evaluated on a real-emission configuration with $n$ extra partons compared to the Born contribution. The phase space integral on the right hand side of \eqn{eq:tsigLO} is of course finite and can be evaluated numerically in four spacetime dimensions.

However, higher-order cross sections are sums of several real-emission and/or virtual contributions that are separately IR divergent and require regularization. For the sake of setting our notation, we recall that at NLO only a single extra emission is allowed, which may be real or virtual. Hence, the full NLO correction reads
\beq
\tsig{NLO}_{ab} = \int_{X+1} \dsig{R}_{ab}\, J_{X+1} 
+ \int_{X} \left(\dsig{V}_{ab} + \dsigk{C}{}{}{}_{ab}\right) J_{X}\,.
\label{eq:tsigNLO}
\eeq
Here $\dsig{R}_{ab}$ and $\dsig{V}_{ab}$ represent the real and virtual cross sections and $\dsigk{C}{}{}{}_{ab}$ denotes the collinear remnant, which accounts for the UV renormalization of the PDFs. Several well-established methods exist to handle the IR singularities in \eqn{eq:tsigNLO}, thus we turn our attention to the NNLO correction immediately.

At NNLO there are precisely two extra emissions which may become unresolved and hence lead to IR singularities. As both of these may be either real or virtual, the complete NNLO correction reads
\beq
\tsigk{NN}{}{LO}{}_{ab} = 
    \int_{X+2} \dsigk{RR}{}{}{}_{ab} J_{X+2}
    +\int_{X+1}\left(\dsigk{R}{}{V}{}_{ab} + \dsigk{C}{1}{}{}_{ab}\right) J_{X+1}
    +\int_{X}\left(\dsigk{}{}{VV}{}_{ab} + \dsigk{C}{2}{}{}_{ab}\right) J_{X}\,.
\label{eq:tsigNkLO}
\eeq
Here $\dsigk{RR}{}{}{}_{ab}$, $\dsigk{R}{}{V}{}_{ab}$ and $\dsigk{}{}{VV}{}_{ab}$ represent the double real, real-virtual and double virtual cross sections, while $\dsigk{C}{1}{}{}_{ab}$ and $\dsigk{C}{2}{}{}_{ab}$ denote the collinear remnants. We recall that the collinear remnants can be written symbolically as~\cite{Ellis:1996mzs}
\beq
\dsigk{C}{1}{}{}_{ab} = \left(\bom{\Gamma}^{(1)} \otimes \dsig{R}\right)_{ab}
\qquad\mbox{and}\qquad
\dsigk{C}{2}{}{}_{ab} = \left(\bom{\Gamma}^{(1)} \otimes \dsig{V}\right)_{ab} + \left(\bom{\Gamma}^{(2)} \otimes \dsig{B}\right)_{ab}
\label{eq:dsigC12}
\eeq
where we define
\beq
\left(\bom{\Gamma}^{(1)} \otimes \dsig{}\right)_{ab} = \Gamma_{1,ac} \otimes \dsig{}_{cb} + \dsig{}_{ac} \otimes \Gamma_{1,cb}
\eeq
and
\beq
\left(\bom{\Gamma}^{(2)} \otimes \dsig{}\right)_{ab} = -\Gamma_{2,ac} \otimes \dsig{}_{cb} - \dsig{}_{ac} \otimes \Gamma_{2,cb} - \Gamma_{1,ac} \otimes \dsig{}_{cd} \otimes \Gamma_{1,db}
\eeq
with
\beq
\Gamma_{1,ab} = \frac{\as}{2\pi} \frac{P^{(0)}_{ab}}{\ep}\,,
\qquad
\Gamma_{2,ab} = \left(\frac{\as}{2\pi}\right)^2 \left[\frac{1}{2\ep^2}\left(P^{(0)}_{ac} \otimes P^{(0)}_{cb} + \beta_0 P^{(0)}_{ab}\right) - \frac{1}{2\ep} P^{(1)}_{ab}\right]\,.
\label{eq:DGLAP-P}
\eeq
Here $P_{ab}$ are the standard space-like splitting functions~\cite{Altarelli:1977zs,Curci:1980uw}\footnote{We are using the \msbar{} scheme here. For other collinear factorization schemes, the splitting functions receive scheme-dependent finite corrections.}, $\beta_0$ is the one-loop coefficient of the QCD beta function and the $\otimes$ symbol indicates the standard integral convolution
\beq
[f \otimes g](x) = \int_0^1 \rd y\,\rd z\,\delta(x-y z)f(y)g(z)\,.
\eeq
In order to regularize all IR singularities in \eqn{eq:tsigNkLO}, we employ the CoLoRFulNNLO subtraction scheme and write
\beq
\bsp
&
\tsig{NNLO}_{ab} = 
    \int_{X+2}\left[
    \dsig{RR}_{ab} J_{X+2} - \dsiga{RR}{1}_{ab} J_{X+1} - \dsiga{RR}{2}_{ab} J_{X} + \dsiga{RR}{12}_{ab} J_{X} 
    \right]
\\ &\quad+
    \int_{X+1}\left\{
    \left[\dsig{RV}_{ab} + \dsigk{C}{1}{}{}_{ab} + \int_1 \dsiga{RR}{1}_{ab}\right] J_{X+1}
    -\left[\dsiga{RV}{1}_{ab} + \dsigk{C}{1,}{A}{1}_{ab} + \left(\int_1 \dsiga{RR}{1}_{ab}\right)^{\!{\rm A}_1}\right] J_{X}
    \right\}
\\ &\quad+
    \int_{X} \left\{
    \dsig{VV}_{ab} + \dsigk{C}{2}{}{}_{ab} + \int_2\left[\dsiga{RR}{2}_{ab} - \dsiga{RR}{12}_{ab}\right] + \int_1\left[\dsiga{RV}{1}_{ab} + \dsigk{C}{1,}{A}{1}_{ab}\right] + \int_1\left(\int_1 \dsiga{RR}{1}_{ab}\right)^{\!{\rm A}_1}
    \right\} J_{X}\,.
\esp
\label{eq:tsigNNLO}
\eeq
Here, the various approximate cross sections have the following interpretation:
\begin{itemize}
\item $\dsiga{RR}{1}_{ab}$ approximates the double real emission cross section $\dsig{RR}_{ab}$ in all single unresolved limits.
\item $\dsiga{RR}{2}_{ab}$ approximates the double real emission cross section $\dsig{RR}_{ab}$ in all double unresolved limits.
\item $\dsiga{RR}{12}_{ab}$ approximates $\dsiga{RR}{2}_{ab}$ in all single unresolved limits {\em and} $\dsiga{RR}{1}_{ab}$ in all double unresolved limits.
\item $\dsiga{RV}{1}_{ab}$ approximates the real-virtual cross section $\dsig{RV}_{ab}$ in all single unresolved limits.
\item $\dsigk{C}{1,}{A}{1}_{ab}$ approximates the collinear remnant $\dsigk{C}{1}{}{}_{ab}$ in all single unresolved limits.
\item $\left(\int_1 \dsiga{RR}{1}_{ab}\right)^{\!{\rm A}_1}$ approximates the integrated single unresolved approximate cross section $\int_1 \dsiga{RR}{1}_{ab}$ in all single unresolved limits.
\end{itemize}
With these subtractions, all three lines on the right hand side of \eqn{eq:tsigNNLO} are rendered finite in four dimensions and can be computed numerically. However, in order to apply \eqn{eq:tsigNNLO} in practical calculations, the formal approximate cross sections which appear in them must be explicitly defined. We turn to this issue next.

% Constructing the subtraction terms

\subsection{Constructing the subtraction terms}
\label{ssec:sub-const}

The universal behavior of QCD squared amplitudes as some number of partons becomes unresolved (soft and/or collinear) is described by IR factorization formul\ae. As mentioned in the Introduction, these are completely known up to NNLO, and in some instances beyond and their general form can be described as follows. Let $\SME{ab,X+k-l}{}{(\mom{}_{X+k-l})}_{l\mathrm{-loop}}$ be the full $l$-loop N$^{k}$LO correction to the squared matrix element, i.e., the correction with a total of $k$ extra emissions where $(k-l)$ of those are real and $l$ are virtual. Consider now the symbolic operator $\bom{U}_{\!j}$, which takes some $j$-fold unresolved ($j\le k-l$) limit or overlap of limits\footnote{Here by overlap of limits, we simply refer to the subsequent application of limits and two different $j$-fold unresolved limits are considered overlapping precisely when their subsequent application produces a configuration that is also $j$-fold unresolved. Note that the successive application limits can also lead to configurations that are more than $j$-fold unresolved. For example,~consider the single ($j=1$) collinear limits $p_1||p_2$ and $p_2||p_3$. Clearly the successive application of these produces the triple collinear configuration $p_1||p_2||p_3$, in which $j=2$ momenta are unresolved. In such cases, we do not consider the limits as overlapping.} of this squared matrix element. The IR limit formula then takes the following symbolic form
\beq
\bom{U}_{\!j} \SME{ab,X+k-l}{}{(\mom{}_{X+k-l})}_{l\mathrm{-loop}} = \left(\frac{\as}{2\pi}\right)^{j} \sum_{i=0}^{l}\mathrm{Sing}_{j}^{(i)} \times \SME{\ha{a}\ha{b},X+k-l-j}{}{(\momh{}_{X+k-l-j})}_{(l-i)\mathrm{-loop}}\,,
\label{eq:IRlimit}
\eeq
i.e., it is a sum over $i$-loop, $j$-fold unresolved universal singular structures $\mathrm{Sing}_{j}^{(i)}$ multiplied by $(l-i)$-loop factorized matrix elements. Depending on the type of unresolved limit, $\mathrm{Sing}_{j}^{(i)}$ involves Altarelli-Parisi splitting functions and eikonal factors, and their appropriate multi-emission and multi-loop generalizations. Thus, these factors are typically matrices in color and/or spin space, and the product in \eqn{eq:IRlimit} above is to be understood accordingly. The factorized matrix element involves $j$ less partons and is evaluated with a corresponding reduced set of momenta $\momh{}$ (e.g., for a soft limit, $\momh{}$ is obtained from $\mom{}$ by simply dropping the soft momenta). The subscripts $\ha{a}$ and $\ha{b}$ indicate that the parton flavors in the factorized matrix element may also differ from the original ones. We note in passing that up to NNLO, the singular factors $\mathrm{Sing}_{j}^{(i)}$ are known to be {\em universal}, i.e., they do not depend on the process under consideration. This makes the construction of a general NNLO subtraction scheme feasible. However, a violation of strict process-independent factorization as implied by \eqn{eq:IRlimit} is possible for initial-state collinear radiation beyond NNLO accuracy~\cite{Catani:2011st}. Any calculational method beyond NNLO will have to address this issue, but it is clearly irrelevant for our present considerations.

In the CoLoRFulNNLO method, these IR limit formul\ae\ are employed as building blocks to construct the approximate cross sections introduced in the previous section. However, they cannot be used directly as subtraction terms for two reasons. First, at any given order, the unresolved regions in phase space overlap, thus care must be taken to avoid multiple subtraction in overlapping regions. Second, these formul\ae\ are only well-defined in the strict IR limits, and as such their definitions must be carefully extended over the full phase space away from the limits.

The issue of overlapping singularities can be addressed simply by the application of the {\em inclusion-exclusion principle}: we must subtract each limit once, then add back the pairwise overlaps of limits, subtract the triple overlaps and so on. Then in order to obtain counterterms that are well-defined over all of phase space, two additional steps must be taken. First, one must specify precisely the momenta entering the factorized matrix elements in the various IR factorization formul\ae. This requires that we specify mappings of sets of momenta $\mom{}_{X+k-l}$ (at NNLO $k=2$, while $l=0$ for the double real correction and $l=1$ for the real-virtual piece) to sets of momenta $\momt{}_{X+k-l-j}$ (where $j \le k-l$, i.e., $j=1$ or $j=2$ for double real emission and $j=1$ for real-virtual emission) which respect momentum conservation and preserve the mass-shell conditions.\footnote{Notice that the reduced set of momenta $\momh{}$ that appear in \eqn{eq:IRlimit} do not necessarily satisfy these conditions. For example, for a soft limit $\momh{}$ only conserves overall momentum in the precise limit when all soft momenta are strictly zero.} Second, the various quantities entering the singular structures in the factorization formul\ae, such as collinear momentum fractions and transverse momenta for collinear splitting and eikonal factors for soft emission, must be precisely defined as functions of the original momenta of the event. After these definitions are fixed, the IR limit formula in \eqn{eq:IRlimit} can be promoted to a (sum of) true subtraction term(s) that is unambiguously defined in any point in phase space,
\beq
\bom{U}_{\!j} \SME{ab,X+k-l}{}{(\mom{}_{X+k-l})}_{l\mathrm{-loop}} \to 
 \sum_{i=0}^{l} {\cal U}_{j}^{(i,l-i)}\,.
\label{eq:IRct}
\eeq
Here
\beq
{\cal U}_{j}^{(i,l-i)} = \left(\frac{\as}{2\pi}\right)^{j} \widetilde{\mathrm{Sing}}_{j}^{(i)} \times \SME{\ti{a}\ti{b},X+k-l-j}{}{(\momt{}_{X+k-l-j})}_{(l-i)\mathrm{-loop}}
\label{eq:IRct1}
\eeq
and $\widetilde{\mathrm{Sing}}_{j}^{(i)}$ simply stands for the expression of the corresponding singular structure incorporating the precise definitions of momentum fractions, eikonal factors and so on. Notice also that the matrix element in \eqn{eq:IRct1} is evaluated over the set of mapped momenta, $\momt{}_{X+k-l-j}$. Obviously, the momentum mappings and definitions of momentum fractions, etc., must be chosen such that they respect the structure of cancellations in all overlapping limits, which is a constraint for the entire construction.

The previously described procedure has been used to construct subtraction terms for handling final-state singularities, whose detailed definitions are available in~\refrs{Somogyi:2006da,Somogyi:2006db,Somogyi:2020mmk}. To extend this approach to hadron-initiated processes, additional subtraction terms are needed to regularize initial-state divergences. Although providing the exact definitions for these new counterterms is beyond the scope of the present paper, we would like to emphasize that a naive extension of the CoLoRFulNNLO method is not possible. Momentum mappings and subtraction terms for initial-state singularities cannot be obtained by simply applying a crossing transformation on their final-state counterparts for several reasons; for instance:
\begin{itemize}
\item The momentum mappings used in~\refrs{Somogyi:2006da,Somogyi:2006db,Somogyi:2020mmk} were designed to keep initial-state momenta unchanged, making them by definition unsuitable for describing collinear emissions from initial-state partons. Hence, we must consider momentum mappings appropriate to this case. We find that we can define all new subtraction terms using 5 elementary mappings, each describing a basic single or double unresolved configuration: i) final-state single collinear; ii) initial-state single collinear; iii) final-state single soft; iv) initial-state double collinear and v) final-state double soft. Some of the required mappings are new (e.g., the final-state single collinear), while others were available in the literature~\cite{Daleo:2006xa,DelDuca:2019ctm}. The iterated single unresolved subtraction terms that appear in $\dsiga{RR}{12}_{ab}$ use compositions of two single unresolved mappings. Also, subtraction terms that describe the overlaps of limits do not require dedicated mappings, but can use one of the elementary mappings for the limits whose overlap they describe.

\item Defining momentum fractions through crossing can lead to spurious singularities. For example, the momentum fraction for the initial-state parton in triple collinear splitting obtained via crossing will vanish at regular points (i.e., points that do not correspond to any IR limit) inside phase
space. But this momentum fraction appears in the denominator of the triple collinear splitting function, leading to spurious singularities. In order to overcome this issue, we introduce appropriate damping functions around these singularities. These damping functions must be constructed such that they do not interfere with the delicate structure of internal cancellations in physical unresolved limits.

\item As an additional complication, integrated initial-state subtraction terms are not simple functions of kinematic invariants, contrary to those for pure final-state radiation~\cite{DelDuca:2015zqa,DelDuca:2016csb,DelDuca:2016ily,Somogyi:2020mmk}. Rather, they are {\em distributions} in the momentum fractions of the initial-state partons. Hence, the appropriate distributional expansions of the integrated counterterms must also be computed.
\end{itemize}

The full details of the extension of the CoLoRFulNNLO method to hadronic collisions will be presented in upcoming publications and here we limit ourselves to giving a symbolic overview of the results.
To begin, we recall that each approximate cross section in \eqn{eq:tsigNNLO} has a clear interpretation in terms of the types of limits it is meant to regularize (e.g., tree-level single and double unresolved limits for $\dsiga{RR}{1}_{ab}$ and $\dsiga{RR}{2}_{ab}$, one-loop single unresolved limits for $\dsiga{RV}{1}_{ab}$, etc.). Hence, approximate cross sections are first of all constructed as sums over the appropriate types of limits. In this regard, we recall that overlaps of limits are considered on the same footing as direct limits: they too must be enumerated and included as per the inclusion-exclusion principle in order to avoid multiple subtraction in overlapping singular regions. Then, for a given physical process, each type of limit can occur several times (e.g., for processes with multiple gluons in the final state, the single soft limit clearly arises separately for each gluon), so all specific singular configurations associated to each type of limit must be accounted for. Thus, approximate cross sections are finally sums of subtraction terms, introduced in \eqn{eq:IRct1}, that each correspond to one specific singular configuration of one specific type of limit. Note however that, since subtraction terms that correspond to overlaps of limits do not use unique momentum mappings, the number of distinct mapped momentum configurations can be less than the total number of subtraction terms. In the following, we refer to each sum of subtractions that share the same mapped momenta as a {\em counter-event}. 

In the remainder of this section, we give explicitly the form of each approximate cross section as a sum of subtraction terms, concentrating on the case of color-singlet production in hadron collisions. This allows for a number of simplifications in the sense that some types of subtraction terms will be absent. However, all the formul\ae\ that we have computed for the colors-singlet case remain valid for more general processes when QCD final state radiation is present already at Born level. In that case, the subtraction terms presented here will simply have to be supplemented by additional ones regularizing the unresolved configurations which do not occur for color-singlet production.

To begin, consider the approximate cross section $\dsiga{RR}{1}_{ab}$, which regularizes the single unresolved limits of the NNLO double real contribution. We write symbolically
\beq
\dsiga{RR}{1}_{ab} = 
	\PS{X+2}(\mom{}_{X+2}) \cA_{1}^{(0)}\,,
\label{eq:dsigRRA1}
\eeq
where the approximation to the matrix element, $\cA_{1}^{(0)}$, is obtained by summing all single soft ($p_r \to 0$) and single collinear ($p_n || p_r$) subtraction terms and subtracting the overlaps as discussed above,
\beq
\cA_{1}^{(0)} =
	\sum_{r \in F}\bigg[
	\cS{r}{(0,0)} 
	+ \sum_{\substack{i \in F \\ i \ne r}} \bigg(\frac12 \cC{ir}{FF (0,0)} - \cC{ir}{FF}\cS{r}{(0,0)}\bigg)
	+ \sum_{c \in I} \bigg(\cC{cr}{IF (0,0)} - \cC{cr}{IF}\cS{r}{(0,0)}\bigg)\bigg]\,.
\label{eq:A1}
\eeq
Here $I$ and $F$ denote the sets of initial-state and final-state partons and the various subtraction terms appearing on the right hand side are explicit realizations of the generic formula for ${\cal U}_{j}^{(i,l-i)}$ in \eqn{eq:IRct1} with $j=1$, $i=0$ and $l=0$. As stated above, each term in \eqn{eq:A1} is defined precisely as a function of the original set of double real momenta $\mom{}_{X+2}$, and the individual terms have the following physical origin: $\cS{r}{(0,0)}$ denotes the subtraction term that regularizes the emission of a single soft gluon, $\cC{ir}{FF(0,0)}$ and $\cC{cr}{IF(0,0)}$ denote subtraction terms regularizing final-final and initial-final collinear singularities, while $\cC{ir}{FF}\cS{r}{(0,0)}$ and $\cC{cr}{IF}\cS{r}{(0,0)}$ account for the double subtraction in the overlapping collinear-soft regions. Note that the factor of $\frac12$ in front of $\cC{ir}{FF(0,0)}$ simply accounts for the fact that the double summation in $i,r \in F$ counts this term twice. The superscript $(0,0)$ signals that these subtraction terms originate from IR limit formul\ae\ that involve the product of tree-level singular structures $\mathrm{Sing}_{1}^{(0)}$ multiplied by zero-loop reduced matrix elements\footnote{More precisely the reduced matrix elements have zero-loop corrections as compared to the Born process. Since the Born process may be loop-induced, the reduced matrix elements may not literally be tree-level. For the sake of simplicity though, we will continue to refer to the reduced matrix elements as zero-loop, one-loop, etc., with the understanding that this loop order is to be understood as compared to the Born process.}. Moreover, in contrast to the generic expression in \eqn{eq:IRct1}, the subscripts on the concrete subtraction terms do not simply give the number of unresolved partons ($j=1$), but instead specify the actual limit from which the term derives. For example, a single unresolved collinear limit is identified by the indices of the partons that become collinear and this is what we write for the collinear-type terms in \eqn{eq:A1}.
Clearly, \eqn{eq:A1} follows the structure for a general approximate cross section that we have laid out: it is a sum of 5 distinct types of limits (single soft, final-final single collinear, final-final collinear-soft overlap, initial-final single collinear and initial-final collinear-soft overlap). After accounting for all singular configurations as implied by the summations in \eqn{eq:A1}, these 5 types of limits give rise to a number of subtraction terms depending on the specific process we consider. For example, for $gg\to Xgg$ production, we find a total of 13 distinct subtraction terms: 2 soft subtractions, 1 final-final collinear subtraction\footnote{Note that $\cC{ir}{FF (0,0)} = \cC{ri}{FF (0,0)}$ and so this term is counted here only once.}, 2 final-final collinear-soft overlaps, 4 initial-final collinear subtractions and finally 4 initial-final collinear-soft overlaps. We note also that the 13 subtraction terms lead to just 7 counter-events, since, as discussed above, the 6 terms that correspond to overlaps do not use separate momentum mappings but rather employ the single soft one. We provide similar details for the rest of the approximate cross sections to be introduced below as well, giving directly the number of types of limits, as well as numbers of subtraction terms and counter-events for the tree-level $gg\to Xgg$ process and the one-loop $gg\to Xg$ process in~\tab{tab:counts}.
\begin{table}
\setlength{\tabcolsep}{12pt}
\renewcommand{\arraystretch}{1.5}
\centering
\begin{tabular}{|c|c|c|c|}
\hline
\parbox{\widthof{cross section}}{\centering ~\\ Approximate cross section \\[-0.5em]~}
& \parbox{\widthof{Nr.\  of types of}}{\centering Nr.\  of types of limits}
& \parbox{\widthof{Nr.\  of subtraction}}{\centering Nr.\  of subtraction terms}
& \parbox{\widthof{Nr.\  of counter-}}{\centering Nr.\  of counter-\\events} \\ 
\hline\hline
\multicolumn{4}{|c|}{Double real} \\ 
\hline
$\cA_{1}^{(0)}$ & 5 & 13 & 7 \\
\hline
$\cA_{2}^{(0)}$ & 5 & 9 & 3 \\
\hline    
$\cA_{12}^{(0)}$ & 12 & 39 & 17 \\
\hline
{\bf Total} & {\bf 22} & {\bf 61} & {\bf 27} \\
\hline\hline
\multicolumn{4}{|c|}{Real-virtual} \\ 
\hline
$\cA_{1}^{(1)}$ & 4 & 7 & 3 \\
\hline
$\cA^{\bom{\scriptstyle \Gamma}}_{1}$ & 1 & 2 & 2 \\
\hline    
$\cA^{\bom{\scriptstyle I}}_1$ & 3 & 5 & 3 \\
\hline
{\bf Total} & {\bf 8} & {\bf 14} & {\bf 3} \\
\hline
\end{tabular}
\caption{\label{tab:counts}
Numbers of types of limits, distinct subtraction terms and counter-events for each approximate cross section for the tree-level process $gg\to Xgg$ (double real) and the 1-loop process $gg\to Hg$ (real-virtual). The total number of counter-events for the real-virtual contribution counts only the independent ones.}
\end{table}

Next, we turn to the double unresolved subtraction terms to double real emission. We write $\dsiga{RR}{2}_{ab}$ symbolically as
\beq
\dsiga{RR}{2}_{ab} = 
	\PS{X+2}(\mom{}_{X+2}) \cA_{2}^{(0)}\,.
\label{eq:dsigRRA2}
\eeq
In this case, one must account for four basic types of limits: double soft ($p_r \to 0$, $p_s \to 0$) , soft-collinear ($p_n || p_r$, $p_s \to 0)$, triple collinear ($p_n || p_r || p_s$) and double collinear ($p_n || p_r$, $p_m || p_s$) and their various overlaps. Concentrating on color-singlet production in hadron collisions we find
\beq
\cA_{2}^{(0)} = 
	\frac{1}{2} \sum_{r \in F}\sum_{\substack{s \in F \\ s \ne r}}\bigg\{
	\cS{rs}{(0,0)} +
	\sum_{c \in I}\bigg[
	\cC{crs}{IFF (0,0)} 
	- \cC{crs}{IFF}\cS{rs}{(0,0)}
    + \sum_{\substack{d \in I \\ d \ne c}}\bigg(
	\cC{cr,ds}{IF,IF (0,0)}	
	- \cC{cr,ds}{IF,IF} \cS{rs}{(0,0)}\bigg)
	\bigg]\bigg\}\,,
\label{eq:A2red}
\eeq
where the various counterterms correspond to the limits implied by the notation. Notice that the repeated application of a triple collinear and double collinear limit leads to a configuration where at least three partons are unresolved, so the corresponding overlapping terms do not appear in \eqn{eq:A2red}. Moreover, as will be shown in an upcoming publication, all soft-collinear type terms cancel among each other for color-singlet production due to the precise definitions we adopt for the subtraction terms. As such, \eqn{eq:A2red} is also free of any terms involving the soft-collinear limit.

Finally, we must consider the overlaps of single and double unresolved regions. In order to avoid double counting in these limits, we introduce the approximate cross section $\dsiga{RR}{12}_{ab}$. This can be written symbolically as
\beq
\dsiga{RR}{12}_{ab} = 
	\PS{X+2}(\mom{}_{X+2}) \cA_{12}^{(0)}\,,
\label{eq:dsigRRA12}
\eeq
where
\beq
\cA_{12}^{(0)} =
	\sum_{s \in F}\bigg[
	\cA_{2}^{(0)}\, \cS{s}{} 
	+ \sum_{\substack{r \in F \\ r \ne s}} \bigg(\frac12 \cA_{2}^{(0)}\, \cC{rs}{FF}
	- \cA_{2}^{(0)}\, \cC{rs}{FF}\cS{s}{}\bigg)
	+ \sum_{c \in I} \bigg( \cA_{2}^{(0)}\, \cC{cs}{IF}
	- \cA_{2}^{(0)}\, \cC{cs}{IF}\cS{s}{} \bigg)\bigg]\,,
\label{eq:A12}
\eeq
with
\bal
\cA_{2}^{(0)}\, \cS{s}{} &=
	\sum_{\substack{r \in F \\ r \ne s}}\bigg[
	\cS{rs}{(0,0)}\, \cS{s}{}
	+ \sum_{c \in I}\bigg(	
	\cC{crs}{IFF (0,0)}\, \cS{s}{}	
	- \cC{crs}{IFF}\cS{rs}{(0,0)}\, \cS{s}{}
	\bigg)
	\bigg]	
	\,, \label{eq:SsA2new}
\\
\cA_{2}^{(0)}\, \cC{rs}{FF} &=
	\cS{rs}{(0,0)}\, \cC{rs}{FF}
	+ \sum_{c \in I}\bigg(
	\cC{crs}{IFF (0,0)}\, \cC{rs}{FF}
	-\cC{crs}{IFF}\cS{rs}{(0,0)}\, \cC{rs}{FF}
	\bigg)
	\,, \label{eq:CrsA2}
\\
\cA_{2}^{(0)}\, \cC{rs}{FF}\cS{s}{} &=
    \sum_{c \in I}
	\cC{crs}{IFF (0,0)}\, \cC{rs}{FF}\cS{s}{}
	\,, \label{eq:CrsFFSsA2new}  
\\
\cA_{2}^{(0)}\, \cC{cs}{IF}  &=
	\sum_{\substack{r \in F \\ r \ne s}} \bigg( \cC{csr}{IFF (0,0)}\, \cC{cs}{IF}
	+ \sum_{\substack{d \in I \\ d \ne c}}
	\cC{cs,dr}{IF,IF (0,0)}\, \cC{cs}{IF}
	 \bigg)
	\,, \label{eq:CasA2new}
\\
\cA_{2}^{(0)}\, \cC{cs}{IF}\cS{s}{} &=
	\sum_{\substack{r \in F \\ r \ne s}}\bigg(
	\cS{rs}{(0,0)}\, \cC{cs}{IF}\cS{s}{}
	+ \cC{csr}{IFF (0,0)}\,  \cC{cs}{IF}\cS{s}{}
	- \cC{csr}{IFF}\cS{rs}{(0,0)}\, \cC{cs}{IF}\cS{s}{}
	\bigg)
	\,.
\label{eq:CasIFSsA2new}
\eal
When writing \eqnss{eq:SsA2new}{eq:CasIFSsA2new}, we have made use of various cancellations at the level of iterated IR factorization formul\ae, as well as some cancellations which occur for color-singlet production due to the specific definitions of the subtraction terms we adopt. The details of these cancellations will be given elsewhere.

Turning to the remaining subtraction terms that regularize real-virtual emission in the second line of \eqn{eq:tsigNNLO}, let us address $\dsiga{RV}{1}_{ab}$ first. This term approximates the real-virtual cross section $\dsig{RV}_{ab}$ in single unresolved limits and can be written symbolically as
\beq
\dsiga{RV}{1}_{ab} = 
	\PS{X+1}(\mom{}_{X+1}) \cA_{1}^{(1)}\,,
\label{eq:dsigRVA1}
\eeq
where for color-singlet production we have
\beq
\cA_{1}^{(1)} =
	\sum_{r \in F}\bigg[
	%\cS{r}{(0,1)} 
%	+ 
    \sum_{c \in I} %\bigg(
    \cC{cr}{IF (0,1)}% - \cC{cr}{IF}\cS{r}{(0,1)}\bigg)
    + \cS{r}{(1,0)} 
	+ \sum_{c \in I} \bigg(\cC{cr}{IF (1,0)} - \cC{cr}{IF}\cS{r}{(1,0)}\bigg)\bigg]\,.
\label{eq:A11}
\eeq
The structure of this subtraction term is similar to that of the single unresolved subtraction term for double real emission in \eqn{eq:A1}. However, given that for color-singlet production the real-virtual term has only a single parton in the final state, the final-final collinear subtraction terms are missing here. Also, the various counterterms in \eqn{eq:A11} are constructed starting from the IR factorization formul\ae\ for one-loop squared matrix elements. As can be seen from \eqn{eq:IRlimit} (the case of a single unresolved limit of the NNLO real-virtual correction is obtained by setting $j=1$, $k=2$ and $l=1$), these formul\ae\ are sums of two terms. Thus, contributions involving tree-level singular structures multiplying one-loop reduced matrix elements appear as counterterms with superscript $(0,1)$, while contributions with one-loop singular structures multiplying zero-loop reduced matrix elements enter the counterterms with superscript $(1,0)$. 
Moreover, we find that for color-singlet production the soft subtraction $\cS{r}{(0,1)}$ is exactly canceled by the sum of collinear-soft overlaps $\cC{cr}{IF}\cS{r}{(0,1)}$, and we have exploited this fact to simplify \eqn{eq:A11}. 

Next, let us consider $\dsigk{C}{1,}{A}{1}_{ab}$, the single unresolved subtraction term to the collinear remnant $\dsigk{C}{1}{}{}_{ab}$. The form of the collinear remnant was given in \eqn{eq:dsigC12}. In particular, notice that it involves the convolution of the $P^{(0)}_{ab}$ splitting functions and the single real emission correction to the process under consideration, $\dsig{R}_{ab}$, which develops IR divergences in the single unresolved regions of phase space. The approximate cross section can then be symbolically written as
\beq
\dsigk{C}{1,}{A}{1}_{ab} = 
    \PS{X+1}(\mom{}_{X+1}) \cA^{\bom{\scriptstyle \Gamma}}_{1}\,.
\label{eq:dsigC1A1}
\eeq
For color-singlet production, this correction again involves just a single parton in the final state, so as for the real-virtual contribution, only the soft and initial-final collinear limits need to be considered. Moreover, it is possible to define the subtraction terms such that for color-singlet production, the soft subtraction, $\cS{r}{(\bom{\scriptstyle \Gamma}\otimes 0,0)}$, is exactly canceled by the sum of the soft-collinear overlaps, $\cC{cr}{IF}\cS{r}{(\bom{\scriptstyle \Gamma}\otimes 0,0)}$. Thus the complete subtraction term reduces to the initial-final collinear contribution and we find
\beq
\cA^{\bom{\scriptstyle \Gamma}}_{1} = 
	\sum_{r \in F}\sum_{c \in I} \cC{cr}{IF (\bom{\scriptstyle \Gamma}\otimes 0,0)}\,.
\label{eq:A1P}
\eeq
The superscript $(\bom{\Gamma}\otimes 0,0)$ refers to the fact that the IR limit formul\ae\ from which the counterterms are constructed involve convolutions of $\bom{\Gamma}$ with tree-level singular structures multiplied by zero-loop reduced matrix elements. In principle, the order of the $\bom{\Gamma}$ operator should also be indicated in the notation for the counterterms. However, up to NNLO these are the only counterterms of this type, hence in order to ease an already elaborate notation, we do not show this order explicitly. 

Last, $\left(\int_1 \dsiga{RR}{1}_{ab}\right)^{\!{\rm A}_1}$ is the single unresolved subtraction term to the integrated subtraction term $\int_1 \dsiga{RR}{1}_{ab}$. As we will briefly discuss below, this in turn can be written symbolically as
\beq
\int_1 \dsiga{RR}{1}_{ab} = \left(\bI^{(0)}_{1}(\ep) \otimes \dsig{R}\right)_{ab}\,,
\label{eq:IntRRA1}
\eeq
where the $\bI^{(0)}_{1}(\ep)$ operator arises after collecting the integrated forms of all subtraction terms in \eqn{eq:A1}. Its precise form will be discussed elsewhere. The corresponding approximate cross section can then be written symbolically as 
\beq
\left(\int_1 \dsiga{RR}{1}_{ab}\right)^{\!{\rm A}_1} = 
	\PS{X+1}(\mom{}_{X+1}) \cA^{\bom{\scriptstyle I}}_1\,,
\label{eq:dsigI1A1}
\eeq
with
\beq
\cA^{\bom{\scriptstyle I}}_1 = 
	\sum_{r \in F}\bigg[
	\cS{r}{(\bom{\scriptstyle I}\otimes 0,0)} 
	+ \sum_{c \in I} \bigg(\cC{cr}{IF (\bom{\scriptstyle I}\otimes 0,0)} - \cC{cr}{IF}\cS{r}{(\bom{\scriptstyle I}\otimes 0,0)}\bigg)\bigg]\,.
\label{eq:A1I}
\eeq
Once more, we have exploited the fact that for color-singlet production, only soft and initial-final collinear limits are relevant. The superscript $(\bom{I}\otimes 0,0)$ implies that the IR limit formul\ae\ from which the counterterms are built involve convolutions of $\bom{I}$ with tree-level singular structures multiplied by zero-loop reduced matrix elements. Again, as was the case with \eqn{eq:A1P}, the notation for the counterterm should include the information carried by the indices of the $\bI$ operator, but to lighten the notation, we do not show these indices. Finally, we note that the existence of universal limits for the integrated approximate cross section in \eqn{eq:IntRRA1} is not guaranteed by QCD factorization formul\ae\ and depends also on the specific definitions of the subtraction terms that we adopt. 

We finish this section by reiterating that all subtraction terms introduced in \eqnss{eq:A1}{eq:A1I} are precisely defined as functions of the original momenta in the double real emission phase space. These definitions will be spelled out in detail in an upcoming publication. Furthermore, since the construction of each subtraction term is based on the appropriate IR factorization formula, all spin correlations in gluon splitting are correctly taken into account.\footnote{In principle, all color correlations are also taken into account, but for color-singlet production, the factorized matrix element has at most three hard partons, so color correlations always reduce to simple multiplication by appropriate linear combinations of squared color-charge operators.} Thus, it is possible to test the cancellation of IR singularities point-wise in phase space by generating sequences of momenta approaching any given IR limit.

% Integrating the subtraction terms

\subsection{Integrating the subtraction terms}
\label{ssec:sub-int}

In order to finish the definition of the subtraction scheme, we must compute the integrals of the counterterms over the momenta of unresolved partons. This computation can be performed once and for all, and we have integrated all counterterms introduced above individually, obtaining fully analytic expressions. Owing to the universal nature of our subtractions terms, the results will be useful also for general processes beyond color-singlet production. The integrated approximate cross sections for the generic case can be build from the integrated subtraction terms presented here, supplemented by the integrals of those subtraction terms that correspond to unresolved limits which do not occur for the color-singlet case.

The complete description of the integration procedure will be given elsewhere (see however \refr{VanThurenhout:2024hmd} for a compact overview) and here we limit ourselves to discussing some general features and to presenting the final results. First, we note that all momentum mappings employed to define the subtraction terms lead to an exact factorization of the real emission phase space in terms of a convolution of the reduced phase space of mapped momenta (denoted by tildes) and an integration measure for the unresolved emission. For the case of a $j$-fold unresolved limit, symbolically we have
\beq
\PS{X+k-l}(\mom{}_{X+k-l}) = [\PS{}]_j \otimes \PS{X+k-l-j}(\momt{}_{X+k-l-j})
\eeq
where $[\PS{}]_j$ represents the phase space measure for the $j$ unresolved emissions. For color-singlet production at NNLO, we must consider the cases $k=2$, $l=0,1$ and $j=1,2$ with $j\le k-l$. The factorized matrix elements entering the counterterms are also evaluated over the same reduced phase space, see \eqn{eq:IRct1}. Thus, the integrated subtraction term can be written generically as
\beq
\bsp
\int \PS{X+k-l}(\mom{}_{X+k-l})\, {\cal U}_j^{(i,l-i)} &= \left(\frac{\as}{2\pi}\right)^j \int [\PS{}]_j\, \widetilde{\mathrm{Sing}}_j^{(i)} 
\\& 
\otimes 
\PS{X+k-l-j}(\momt{}_{X+k-l-j}) \SME{\ti{a}\ti{b},X+k-l-j}{}{(\momt{}_{X+k-l-j})}_{(l-i)\mathrm{-loop}}
\\ &=
\left(\frac{\as}{2\pi}\right)^j \left(\int [\PS{}]_j\, \widetilde{\mathrm{Sing}}_j^{(i)} \otimes \dsigk{R}{k-l-j}{V}{l-i}\right)_{ab}\,,
\esp
\label{eq:ICT}
\eeq
where we have introduced the notation $\dsigk{R}{m}{V}{n}$ to denote the $m$-fold real and $n$-fold virtual correction to the differential cross section. In this notation $\dsig{B} = \dsigk{R}{0}{V}{0}$, $\dsig{R} = \dsigk{R}{1}{V}{0}$, $\dsig{V} = \dsigk{R}{0}{V}{1}$ and so on. Then, the integration of the subtraction terms over the measure for unresolved emission can be performed once and for all. Since the phase space factorization is convolutional in nature, the integrated counterterms,
\beq
[U_j^{(i)}] \equiv \int [\PS{}]_j\, \widetilde{\mathrm{Sing}}_j^{(i)}\,,
\eeq
turn out to be linear combinations of Dirac-delta and plus-distributions, as well as regular terms in the convolution variables. 
Details about the integration of all the counterterms will be presented in dedicated publications and here we limit ourselves to summarizing the main features.  All integrated subtraction terms were reduced to a number of master integrals, see \tab{tab:mi_counts}, which were subsequently computed analytically to the required order is $\eps$. In particular, the ones arising from the double unresolved counterterms of \eqn{eq:A2red} were obtained using the method of differential equations~\cite{Kotikov:1990kg,Kotikov:1991pm,Kotikov:1991hm,Gehrmann:1999as}, while for the others we employed direct integration techniques. For the latter we used the method of \refr{Brown:2008um} as described in \refr{Anastasiou:2013srw} and implemented in {\sc PolyLogTools}~\cite{Duhr:2019tlz}. All integrated counterterms can be computed in terms of multiple polylogarithms (MPLs)~\cite{Mpls1} evaluated at algebraic arguments. We note that, for generic values of the momentum fractions $x_a$ and $x_b$ of the initial-state partons, we only need to evaluate the integrals up to orders in $\eps$ that involve at most MPLs of weight two. Higher orders, which involve MPLs of weight three, only contribute for $x_a=1$ or $x_b=1$, after expanding, e.g., $(1-x_a)^{-1+m\eps}$ into distributions,
\begin{equation}
(1-x_a)^{-1+m\varepsilon} = -\frac{1}{m\varepsilon}\delta(1-x_a) + \ldots\,,
\end{equation}
where the dots represent regular terms (in $\eps$) that contain
plus-distributions.  Similarly, weight-four contributions only arise
for $x_a=x_b=1$, and so they are constant. This has important
practical implications. First, the higher-weight terms have a simpler
functional dependence on $x_a$ and $x_b$, and are thus easier to
evaluate. Second, the only non-constant MPLs have weight at most
three, and it is known that these can always be expressed in terms of
ordinary logarithms and classical
polylogarithms~\cite{Lewin:1981,Goncharov:1996dce,Kellerhals:1995},
\begin{equation}
\label{eq:Li_series}
    \Li_n(x) = \sum_{k=1}^\infty\frac{x^k}{k^n}\,,\qquad |x|<1\,.
\end{equation}
It follows that all our integrated counterterms can be expressed only
in terms of these functions. We have applied the algorithm of
ref.~\cite{Duhr:2011zq} to write our results exclusively in terms of
logarithms, $\Li_2$ and $\Li_3$ with arguments within the unit circle,
so that the series in \eqn{eq:Li_series} is convergent. This
allows us to easily evaluate all special functions appearing in the
integrated counterterms in a fast and stable way. However,
the complete expressions for the finite parts of the insertion
operators are quite elaborate.
\begin{table}
\setlength{\tabcolsep}{12pt}
\renewcommand{\arraystretch}{1.5}
\centering
\begin{tabular}{|c|c|}
\hline
\parbox{\widthof{cross section}}{\centering ~\\ Approximate cross section \\[-0.5em]~}
&  \parbox{\widthof{Nr.\ of master}}{\centering {Nr.\ of master integrals}} \\ 
\hline\hline
\multicolumn{2}{|c|}{Double real} \\ 
\hline
$\cA_{1}^{(0)}$ & 11 \\
\hline
$\cA_{2}^{(0)}$ & 42 \\
\hline    
$\cA_{12}^{(0)}$ & 104 \\
\hline
{\bf Total} & {\bf 157} \\
\hline\hline
\multicolumn{2}{|c|}{Real-virtual} \\ 
\hline
$\cA_{1}^{(1)}$ & 24 \\
\hline
$\cA^{\bom{\scriptstyle \Gamma}}_{1}$ & 10 \\
\hline    
$\cA^{\bom{\scriptstyle I}}_1$ & 65 \\
\hline
{\bf Total} & {\bf 99} \\
\hline
\end{tabular}
\caption{\label{tab:mi_counts}
Number of master integrals that appear in the evaluation of the integrated subtraction terms.}
\end{table}

The various integrated subtraction terms contributing to each approximate cross section can be gathered into a single insertion operator with poles in $\ep$. For example, evaluating the integral of $\dsiga{RR}{1}$, we have $k=2$, $l=0$ and $j=1$ in \eqn{eq:ICT}, which forces $i=0$ since $i\le l$. So symbolically
\beq
\int_1 \dsiga{RR}{1}_{ab} = \frac{\as}{2\pi} \sum_{U}\, \left([U_1^{(0)}] \otimes \dsigk{R}{1}{V}{0}\right)_{ab}
= \left(\bI^{(0)}_{1}(\ep) \otimes \dsig{R}\right)_{ab}\,,
\label{eq:I10def}
\eeq
where we have simply set $\bI^{(0)}_{1}(\ep) = \frac{\as}{2\pi} \sum_{U}\, [U_1^{(0)}]$ to recover the expression in \eqn{eq:IntRRA1}. As noted below \eqn{eq:IRlimit}, in IR factorization formul\ae\, the parton flavors of the factorized matrix elements may differ from the original ones. This property then holds also for the subtraction terms and their integrated forms. Thus, in addition to acting on the momentum fractions in $\dsig{R}$ via the integral convolution, the operator  $\bI^{(0)}_{1}(\ep)$  also acts on the parton flavors of the real emission cross section,
\beq
\left(\bI^{(0)}_{1}(\ep) \otimes \dsig{R}\right)_{ab} = 
\sum_{c,d} \bI^{(0)}_{1,ac,bd}(\ep) \otimes \dsig{R}_{cd}\,.
\eeq
All other insertion operators to be introduced in the following share this structure, although, as in \eqn{eq:I10def}, we lighten the notation by not indicating the flavor indices of insertion operators and the corresponding flavor summations explicitly. After adding this integrated approximate cross section to the real-virtual contribution (including the appropriate collinear remnant), we find that the $\ep$-poles cancel, which is a check on the correctness of the scheme.

Turning to the two other approximate cross sections that enter the regularization of double real emission, $\dsiga{RR}{2}_{ab}$ and  $\dsiga{RR}{12}_{ab}$, we find
\beq
\int_2 \dsiga{RR}{2}_{ab} = \left(\bI^{(0)}_{2}(\ep) \otimes \dsig{B}\right)_{ab}
\qquad\mbox{and}\qquad
\int_2 \dsiga{RR}{12}_{ab} = \left(\bI^{(0)}_{12}(\ep) \otimes \dsig{B}\right)_{ab}\,.
\label{eq:IntRRA2-IntRRA12}
\eeq
Obviously, the cross sections appearing on the right hand sides are the Born cross sections, since all unresolved radiation has been integrated out.

Next, consider the integrated version of $\dsiga{RV}{1}_{ab}$, i.e., the single unresolved approximation to the real-virtual contribution. As discussed below \eqn{eq:A11}, the full subtraction term is a sum of two contributions, one involving tree-level singular structures and one-loop reduced matrix elements and the other one-loop singular structures and zero-loop reduced matrix elements. This structure is then inherited by the integrated approximate cross section which can be written as
\beq
\int_1 \dsiga{RV}{1}_{ab} = \left(\bI^{(0)}_{1}(\ep) \otimes \dsig{V} + \bI^{(1)}_{1}(\ep) \otimes \dsig{B}\right)_{ab}\,.
\label{eq:IntRVA1}
\eeq
Clearly, this structure also follows formally from \eqn{eq:ICT} with $k=1$, $l=1$ and $j=1$, since now both $i=0$ and $i=1$ contributions are allowed.

The remaining two approximate cross sections, $\dsigk{C}{1}{}{}_{ab}$ and $\left(\int_1 \dsiga{RR}{1}_{ab}\right)^{\!{\rm A}_1}$, both involve only the Born matrix element, and their integrated versions may be written in the form
\beq
\int_1 \dsigk{C}{1}{,A}{1}_{ab} = \left(\bI^{(0,0)}_{\Gamma,1}(\ep) \otimes \dsig{B}\right)_{ab}
\qquad\mbox{and}\qquad
\int_1 \left(\int_1 \dsiga{RR}{1}_{ab}\right)^{\!{\rm A}_1} = \left(\bI^{(0,0)}_{1,1}(\ep) \otimes \dsig{B}\right)_{ab}\,.
\eeq

Having computed the integrated forms of all subtraction terms, we can now combine the results with the virtual matrix elements and collinear remnants. After this combination, the integrand on the third line of \eqn{eq:tsigNNLO} becomes 
\beq
\bsp
&
\dsig{VV}_{ab} + \dsigk{C}{2}{}{}_{ab} + \int_2\left[\dsiga{RR}{2}_{ab} - \dsiga{RR}{12}_{ab}\right] + \int_1\left[\dsiga{RV}{1}_{ab} + \dsigk{C}{1,}{A}{1}_{ab}\right] + \int_1\left(\int_1 \dsiga{RR}{1}_{ab}\right)^{\!{\rm A}_1} 
\\&=
\dsig{VV}_{ab} + \dsigk{C}{2}{}{}_{ab} 
+ \left[\bI^{(0)}_{1}(\ep) \otimes \dsig{V} + \left(\bI^{(0)}_{2}(\ep) - \bI^{(0)}_{12}(\ep) + \bI^{(1)}_{1}(\ep) + \bI^{(0,0)}_{\Gamma,1}(\ep) + \bI^{(0,0)}_{1,1}(\ep)\right)\otimes \dsig{B}\right]_{ab}\,.
\esp
\label{eq:IntVVNNLO}
\eeq
The explicit $\eps$-poles present in the various terms in \eqn{eq:IntVVNNLO} then cancel and the complete expression can be evaluated numerically in four dimensions. For the case of Higgs boson production in HEFT with only gluons, all matrix elements are extremely compact and the cancellation of $\eps$-poles can easily be checked analytically. Indeed, using the expressions presented in \appx{appx:poles} (or indeed the known structure of two-loop IR divergences~\cite{Catani:1998bh}), it is straightforward to show that the sum in \eqn{eq:IntVVNNLO} is free of $\ep$-poles, which gives a strong check on the correctness of our calculations.

%%%%%%%%%%%%%%%%%%%%%%%%%%%%%%%%%%%%%%%%%%%%%%%%%%%%%%%%%%%%%%%%%%%
%%%%%%%%%%%%%%%%%%%%%%%%%%%%%%%%%%%%%%%%%%%%%%%%%%%%%%%%%%%%%%%%%%%

%%%
%%% NNLOCAL
%%%

\section{The {\tt NNLOCAL} code}
\label{sec:nnlocal}

In this section, we introduce {\tt NNLOCAL}, a
proof-of-concept parton level Monte Carlo code implementing the
subtraction scheme described above. The code can be obtained at 
\url{https://github.com/nnlocal/nnlocal.git} and some details about 
the installation and running options can be found in 
\appx{app:install}. {\tt NNLOCAL} is written in {\tt
  Fortran77} and its architecture is based on a previous version of
the well-known Monte Carlo program
MCFM~\cite{Campbell:1999ah,Campbell:2011bn,Campbell:2019dru}, to be
more precise we refer here to MCFM-4.0. In particular, phase space
integrations are handled with the {\tt Vegas}
algorithm~\cite{Lepage:1977sw} for adaptive multidimensional Monte
Carlo integration. This algorithm implements adaptive importance
sampling by first building an integration grid which adapts to the
integrand iteratively. After this warm-up stage of grid refinement,
results are collected with a fixed grid and a Monte Carlo estimate of
the integral is computed. Once the {\tt Vegas} grid has been computed,
one can of course use it to accumulate results from many independent
runs executed in parallel. For a more efficient use of computational
resources, in {\tt NNLOCAL} we modified the normal workflow of {\tt
  Vegas} by introducing the possibility to run in parallel also the 
refinement of the grids. For this, we adopted the following
solution. First we run $n$ independent instances of {\tt NNLOCAL}, all
performing a single iteration of grid refinement and we let {\tt
  Vegas} generate $n$ independent versions of the grids. Then, for
each integration variable, $a$, we combine the $n$ versions of the
grid by summing the corresponding cumulative distributions, $c^a_j(x)$, and
dividing the result by $n$.
\begin{equation}
\bar{c}^a(x)=\frac{1}{n}\sum_{j=1}^n\,c^a_j(x)\,.
\end{equation}
Finally, we compute the $N$ new grid points, $x^a_i$, of a regular {\tt Vegas}
grid, by solving the equation
\begin{equation}
\bar{c}(x^a_i)=\frac{i}{N} \quad, \quad i=1,...,N
\end{equation}
where the integer $i$ runs over the $N$ grid divisions. Since the
averaged cumulative is the sum of $n$ linear functions, it is also
linear, and the numerical solution to the equation above is
straightforward to compute with standard routines. The subsequent
parallel iteration is then started with the new grid and the above procedure 
is repeated until the number of requested grid refinement iterations is reached. 
We point out that with such a procedure it is possible to combine results
produced on a computer using any number of cores. Furthermore, in {\tt
  NNLOCAL} we have included support for the visualization of the grids
at every step of the warm-up stage\footnote{In {\tt NNLOCAL}, {\tt gnuplot} scripts
are automatically generated to plot the cumulative of the
distributions associated to the {\tt Vegas} grids.}, which can be useful to
assess the quality of the integration procedure.

\begin{figure}
    \centering
    \includegraphics[width=0.3\linewidth]{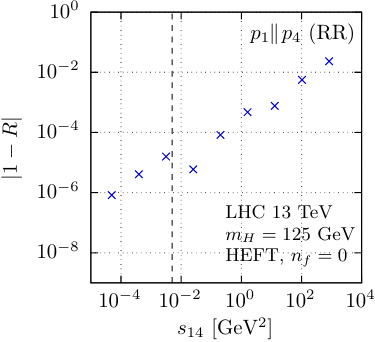}
    \hspace{1.5em}    
    \includegraphics[width=0.3\linewidth]{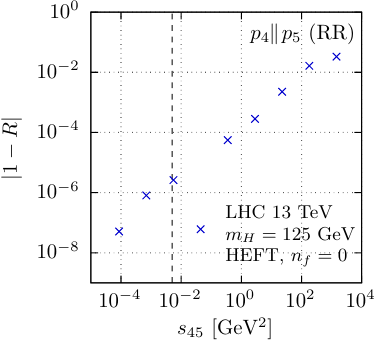}
    \hspace{1.5em}
    \includegraphics[width=0.3\linewidth]{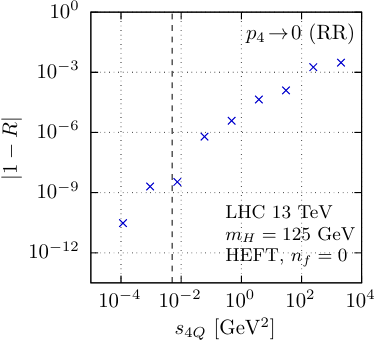}   
    \\[1.5em]
    \includegraphics[width=0.3\linewidth]{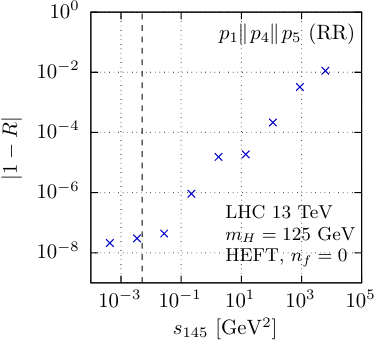}
    \hspace{1.5em}
    \includegraphics[width=0.3\linewidth]{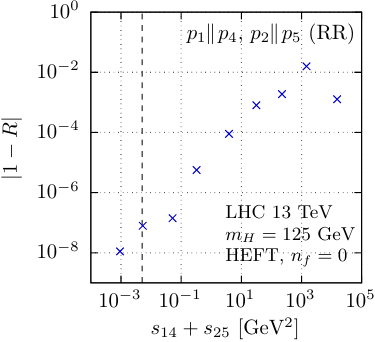}
    \hspace{1.5em}
    \includegraphics[width=0.3\linewidth]{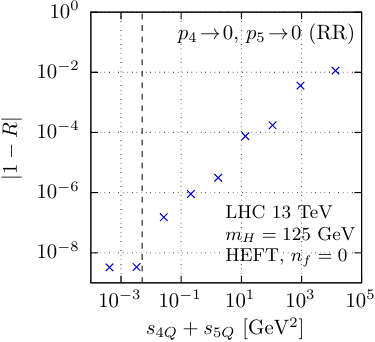}    
    \caption{The behavior of $|1-R|$ in the $g(p_1) + g(p_2) \to H(p_3) + g(p_4) + g(p_5)$ subprocess as various single (first row) and 
    double (second row) unresolved kinematic limits are 
    approached. Here $R$ denotes the ratio 
    $R = (\cA_{1}^{(0)} + \cA_{2}^{(0)} - \cA_{12}^{(0)})/\SME{gg,H+2}{(0)}{}$. 
    The proximity to a given limit is measured by the smallness of an 
    appropriate invariant. The dashed vertical line denotes the default 
    value of the technical cutoff $s_{\mathrm{min}} = 5\cdot 
    10^{-3}$~GeV${}^2$. Large outliers such as the one visible in the 
    $p_4||p_5$ limit are due to large accidental cancellations.
    }
    \label{fig:RRlims}
\end{figure}

As stated before, the code in its current state deals with Higgs boson
production in proton-proton collisions in the HEFT approximation with
no light quarks. Given that {\tt NNLOCAL} is a proof-of-concept code, we supplemented
our tool with a number of facilities to test its behavior. As for the
double real correction (that we dub {\tt real} in {\tt NNLOCAL}), we
have introduced a dedicated phase space generator to probe every
singular region in the phase space starting from a non-singular
configuration and approaching the desired region point-by-point. In
this way it is possible to monitor the level of cancellation among the
matrix element and the sum of all counterterms. 
We demonstrate this by examining the ratio $R$ of the sum of 
all double real subtraction terms $(\cA_{1}^{(0)}+\cA_{2}^{(0)}-\cA_{12}^{(0)})$ and the squared matrix element in \fig{fig:RRlims}. 
There we show $|1-R|$ as a function of the small 
invariant whose vanishing characterizes a given kinematic limit. Decreasing 
the invariant by 7--8 orders of magnitude from its initial value, we observe 
the convergence of this ratio to 1 in all cases as the singularity is 
approached. However, due to finite precision arithmetic, numerical 
instabilities eventually set in as we further decrease the value of the 
invariant, driving the ratio away from 1. 
In order to circumvent this issue, we introduce a technical 
cutoff, $s_{\mathrm{min}}$, and do not allow two-parton invariants 
to take values lower than this.\footnote{Note that even if one were to 
employ arbitrary precision arithmetic, some technical cutoff would still be 
required since both the matrix element and the subtraction terms diverge and 
thus are technically undefined in the exact limits.} We indicate our default 
choice of $s_{\mathrm{min}} = 5\cdot 10^{-3}$~GeV${}^2$ on the plots with 
dashed vertical lines. We will demonstrate below that this choice is small 
enough such that further lowering it does not influence the values of physical 
observables.

\begin{figure}
    \centering
    \includegraphics[width=0.30\linewidth]{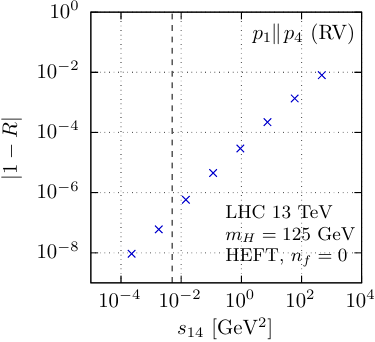}
    \hspace{1.5em}
    \includegraphics[width=0.30\linewidth]{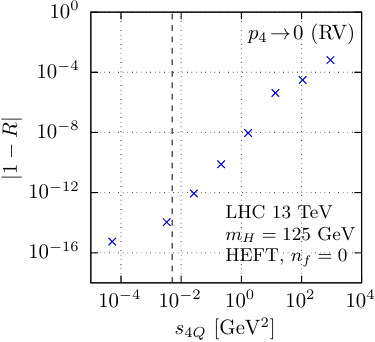}
    \caption{The behavior of $|1-R|$ in the $g(p_1) + g(p_2) \to H(p_3) + g(p_4)$ subprocess at one-loop order as single  unresolved kinematic 
    limits are approached. Here $R$ denotes the ratio 
    $R = (\cA_{1}^{(1)}+\cA^{\bom{\scriptstyle \Gamma}}_{1} +\cA^{\bom{\scriptstyle I}}_{1})/(\SME{gg,H+1}{}{}_{1\mathrm{-loop}} + \bom{\Gamma}^{(1)} \otimes \SME{gg,H+1}{(0)}{} + \bI_1^{(0)} \otimes \SME{gg,H+1}{(0)}{})$.   
    The proximity to a given limit is measured by 
    the smallness of an appropriate invariant.  The dashed vertical line 
    denotes the default value of the technical cutoff $s_{\mathrm{min}} = 5\cdot 
    10^{-3}$~GeV${}^2$.
    }
    \label{fig:RVlims}
\end{figure}

In the real-virtual correction (dubbed {\tt virt} in {\tt NNLOCAL}), we
encounter two types of singularities: explicit $\eps$-poles and phase
space divergences. We have used the setup mentioned above to check
the local cancellation of phase space divergences and this is 
demonstrated in \fig{fig:RVlims}. In this case, we concentrate on the ratio 
of the sum of subtraction terms 
$(\cA_{1}^{(1)}+\cA^{\bom{\scriptstyle \Gamma}}_{1} +\cA^{\bom{\scriptstyle I}}_{1})$ 
and the sum of the real-virtual matrix element, the appropriate collinear 
remnant and the integrated single unresolved subtraction terms. These sums are individually finite in $\eps$ but are divergent as the final-state parton becomes 
unresolved. Once again, we observe the convergence of this 
ratio to 1, which establishes the correct cancellation of kinematic 
singularities. Furthermore in {\tt
  NNLOCAL}, matrix elements and counterterms are all coded as vectors
of coefficients of the corresponding Laurent-expansion in $\eps$, so
that we have direct access to the individual and total poles and so
can explicitly check their numerical cancellation. The last
consideration also applies to the double virtual contributions, that
we dub {\tt born} in our code.

As for the sum of all the integrated counterterms, we notice that in
certain regions, care must be taken to obtain an implementation that
delivers accurate numerical results.  For example, when $x_a = x_b$ (but
also along some other curves), a naive evaluation of certain
integrated subtraction terms produces a result of the form “0/0”,
while in fact the $x_a \to x_b$ limit is well-behaved and
finite. Clearly these cases must be handled to avoid undefined results
and large instabilities. In {\tt NNLOCAL}, we use dynamical switching
to quadruple precision in order to improve stability and eventually
employ a technical cutoff, $\Delta x_{\mathrm{min}}$, to avoid possible 
undefined expressions
whenever such regions are approached. We find that we typically 
need to switch to quadruple precision in about 6\% of the evaluations and 
we choose $\Delta x_{\mathrm{min}} = 10^{-5}$ as the default value of the 
cutoff. The further lowering of $\Delta x_{\mathrm{min}}$ does not influence 
the value of physical quantities as we will demonstrate below.

\begin{figure}
    \centering
    \includegraphics[width=0.4\linewidth]{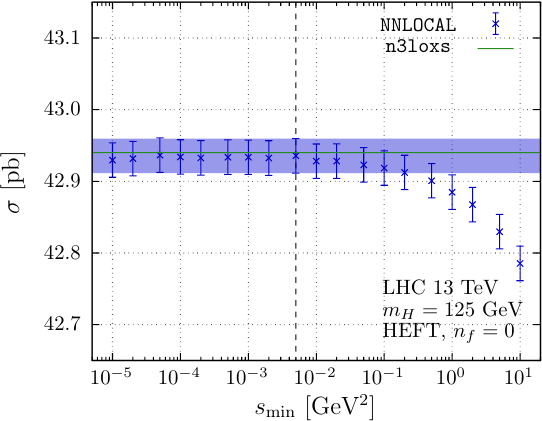}  
    \hspace{1.5em}
    \includegraphics[width=0.4\linewidth]{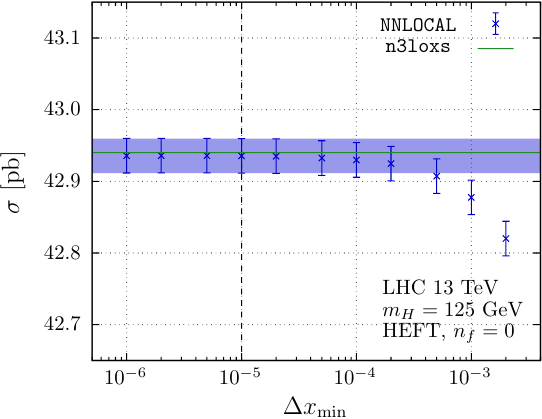}   
    \caption{The dependence of the total cross section on the technical cutoffs 
    $s_{\mathrm{min}}$ (left) and $\Delta x_{\mathrm{min}}$ (right). Dashed 
    vertical lines show the default values of 
    $s_{\mathrm{min}}=5\cdot 10^{-3}$~GeV${}^2$ and 
    $\Delta x_{\mathrm{min}} = 10^{-5}$. The blue band corresponds to 
    the prediction at default cutoff values. The green line represents the prediction from {\tt n3loxs} obtained with the modifications discussed in the text.}
    \label{fig:cuts}
\end{figure}

In order to validate our code, we have computed the total cross
section for the production of a Higgs boson at the LHC with 13~TeV
center of mass energy at NNLO in HEFT without light quarks for several
different values of $m_H$. We then performed a tuned comparison of our
results to the code {\tt n3loxs}~\cite{Baglio:2022wzu}. In order to
synchronize the two codes fully, we made two changes to the publicly
available version of {\tt n3loxs}. First, we imported into {\tt
  n3loxs} the routine for the computation of the strong coupling from
{\tt NNLOCAL}. Second, we excluded quark channels in {\tt n3loxs}. In
the calculations we used the {\tt NNPDF31_nnlo_as_0118} PDF
set~\cite{NNPDF:2014otw} and performed validation runs for several
choices of the renormalization and factorization scales. 
The results for 
the scale choice $\mu_R = \mu_F = m_H$ are given in \tab{tab:xstot}. 
We observe perfect agreement between the results of {\tt n3loxs} and 
{\tt NNLOCAL}, with relative differences in the sub-permille range over 
the full range of Higgs boson masses.
\begin{table}[ht]
\setlength{\tabcolsep}{12pt}
\renewcommand{\arraystretch}{1.5}
    \centering
    \begin{tabular}{|c|c|c|}
    \hline%\hline
         $m_H$ [GeV] &  {\tt n3loxs} (gg) &  {\tt
         NNLOCAL} (gg) \\
    \hline\hline
         100 & $65.72$ pb & $65.74 (4)$ pb \\ 
    \hline
         125 & $42.94$ pb & $42.94 (2)$ pb\\ 
    \hline
         250 & $9.730$ pb & $9.733 (5)$ pb \\ 
    \hline
         500 & $1.626$ pb & $1.626 (1)$ pb \\ 
    \hline
         1000 & $173.7$ fb & $173.7 (1)$ fb \\ 
    \hline
         2000 & $8.794$ fb & $8.790 (5)$ fb \\ 
    \hline%\hline
    \end{tabular}
    \caption{The total NNLO cross section for Higgs boson production at the LHC with 13~TeV center of mass energy in HEFT with $\Nf=0$ light quarks. The errors on the results obtained with {\tt NNLOCAL} represent the estimated uncertainties of the Monte Carlo integrations. The estimated uncertainty of the {\tt n3loxs} result is beyond the last displayed digit in each case. The shown results were obtained on a MacBook Pro laptop with an M2 processor with 8 CPU cores.}
    \label{tab:xstot}
\end{table}
We note furthermore that the obtained values are very stable under the variation of the technical cuts $s_{\mathrm{min}}$ and $\Delta x_{\mathrm{min}}$ introduced above. This is demonstrated in \fig{fig:cuts}. The left panel shows the value of the total cross section obtained for a Higgs boson of mass $m_H=125$~GeV as a function of the value of the smallest allowed two-parton invariant $s_{\mathrm{min}}$. The right panel shows the total cross section as a function of the cutoff parameter $\Delta x_{\mathrm{min}}$ discussed above. In both cases we observe a very stable plateau as the cutoffs are taken to zero.
The runtime per mass value required to obtain the displayed precision in 
\tab{tab:xstot} with {\tt NNLOCAL} is around 20 minutes on a MacBook Pro 
laptop with an M2 processor with 8 CPU cores.

In order to demonstrate the impact of the NNLO correction, in 
\tab{tab:xstot_nnlo} we present the pure NNLO contribution to the total 
cross section, along with its breakdown into the 
double virtual, real-virtual and double real pieces. 
The value of the full NNLO contribution is of course fixed, however the 
values of the individual pieces depend on the specific scheme that is used 
to regularize IR singularities. We observe that in our setup, the bulk of 
the NNLO correction arises from the double virtual piece, while the 
real-virtual and especially the double real pieces are very much 
sub-dominant, the latter contributing only at around the 0.5--2\% level, depending on the mass of the Higgs boson.
\begin{table}[t]
\setlength{\tabcolsep}{12pt}
\renewcommand{\arraystretch}{1.5}
    \centering
    \begin{tabular}{|c|c|c|c|c|}
    \hline%\hline
         $m_H$ [GeV] & NNLO only & double virtual 
         & real-virtual & double real \\
    \hline\hline
         100 & $19.76(5)$ pb & $20.76(5)$ pb & 
         $-1.318(1)$ pb & $0.321(9)$ pb \\ 
    \hline
        125 & $12.64(1)$ pb & $13.16(1)$ pb & 
        $-0.7492(4)$ pb & $0.235(3)$ pb \\ 
    \hline
        250 & $2.724(2)$ pb & $2.781(2)$ pb & 
        $-0.10475(4)$ pb & $0.0474(2)$ pb \\         
    \hline
        500 & $0.4420(3)$ pb & $0.4460(3)$ pb & 
        $-0.009645(4)$ pb & $0.00564(1)$ pb \\         
    \hline
        1000 & $47.23(3)$ fb & $47.30(3)$ fb & 
        $-0.4699(2)$ fb & $0.3982(8)$ fb \\         
    \hline
        2000 & $2.496(2)$ fb & $2.492(2)$ fb & 
        $-0.007597(2)$ fb & $0.01220(3)$ fb \\         
    \hline%\hline
    \end{tabular}
    \caption{The NNLO correction to the total cross section for Higgs boson production at the LHC with 13~TeV center of mass energy in HEFT with $\Nf=0$ light quarks. The breakdown of the result into double virtual, real-virtual and double real pieces is also shown. The errors represent the estimated uncertainties of the Monte Carlo integrations.}
    \label{tab:xstot_nnlo}
\end{table}

\begin{figure}[ht]
   \centering
   \includegraphics[width=0.4\linewidth]{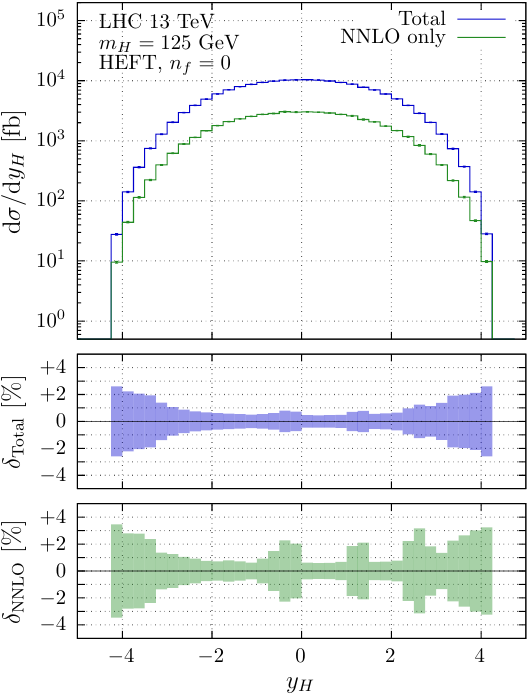} 
   \hspace{1.5em}
   \includegraphics[width=0.4\linewidth]{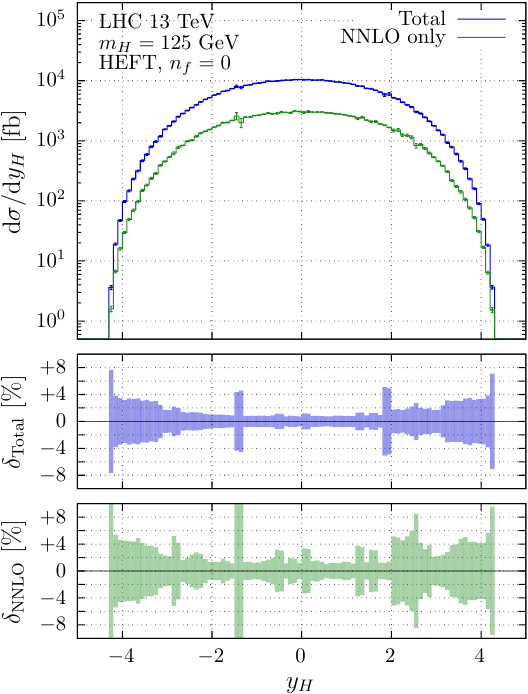}  
    \caption{The rapidity distribution of the Higgs boson at NNLO in HEFT with $\Nf=0$ light quarks (blue). The NNLO contribution is also shown separately (green). Distributions with bin width $\Delta y = 0.25$ (left) and $\Delta y = 0.1$ (right) are shown. The errors represent the estimated Monte Carlo uncertainties. The lower panels in both plots show the relative uncertainties for the total distribution and the NNLO contribution. The shown results were obtained on a MacBook Pro laptop with an M2 processor with 8 CPU cores with a runtime of around 1 hour and 15 minutes.}
    \label{fig:yH}
\end{figure}

Since our code is completely differential in all particle momenta, any infrared and collinear-safe quantity can be computed by simply implementing the proper analysis routine. By way of illustration, we present the rapidity distribution of a Higgs boson of mass $m_H=125$~GeV at the 13~TeV LHC (in HEFT with $\Nf=0$) in \fig{fig:yH}. 
In both plots in the figure, the upper panels show the total distribution as well as the NNLO contribution, while filled bands in the lower panels indicate the relative Monte Carlo error estimates.
The figure was obtained on the same architecture as the numbers in \tabs{tab:xstot}{tab:xstot_nnlo}. The runtime was around 1 hour and 15 minutes.
In the left plot, we present results with a bin width of $\Delta y = 0.25$. We observe very good numerical convergence and stability over the full range of values, spanning about four orders of magnitude, both for the total distribution and for the NNLO contribution. In particular, the total and NNLO distributions generally have an uncertainty of less than 1\% and 2\% over the central rapidity range of $|y_H|<2.5$. In order to demonstrate the numerical stability of our code under more demanding conditions, in the right plot of \fig{fig:yH}, we show the same rapidity distribution, but with bin width of $\Delta y = 0.1$. Overall, we still observe good stability and convergence, however, with this rather fine binning, the appearance of some spikes in the predictions is evident. This phenomenon, commonly dubbed ``misbinning'', is a well-known and more or less unavoidable consequence of using a local subtraction scheme to regularize IR divergences. Due to the binned nature of the distributions,  sometimes momentum mappings cause the weights associated to the matrix element and subtraction to end up in different bins. This leads to an apparent uncanceled singularity as far as the bin is concerned, with the appearance of unphysical spikes in distributions during Monte Carlo integration. Obviously computing with wider bins mitigates the effect of misbinning.

\begin{figure}[ht]
   \centering
   \includegraphics[width=0.4\linewidth]{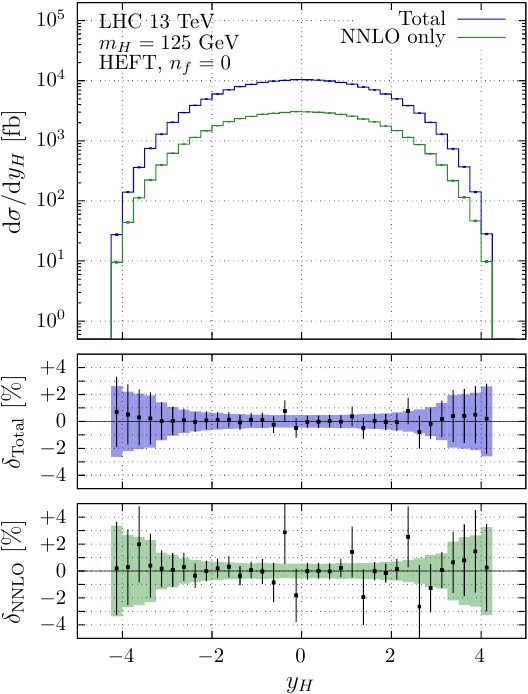}
   \hspace{1.5em}
   \includegraphics[width=0.4\linewidth]{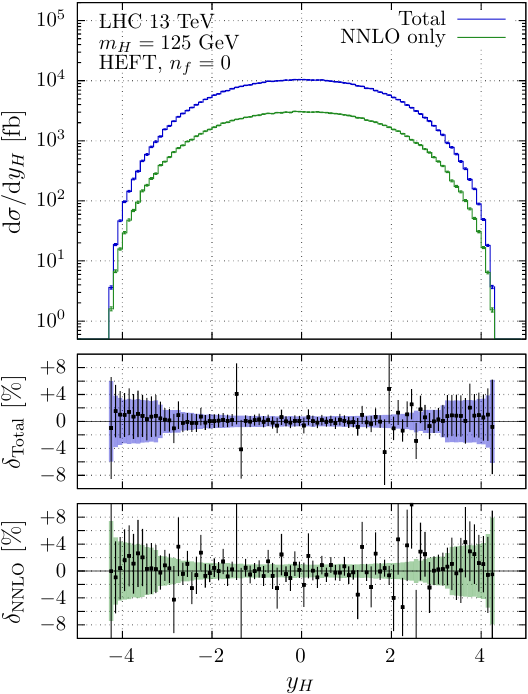}
    \caption{The trimmed rapidity distribution of the Higgs boson at NNLO in HEFT with $\Nf=0$ light quarks (blue). The NNLO contribution is also shown separately (green). Distributions with bin width $\Delta y = 0.25$ (left) and $\Delta y = 0.1$ (right) are shown. The errors represent the estimated Monte Carlo uncertainties. The lower panels in both plots show the relative uncertainties for the total distribution and the NNLO contribution. The black data points in the lower panels show the ratio of the untrimmed and trimmed results bin by bin. The shown results were obtained on a MacBook Pro laptop with an M2 processor with 8 CPU cores with a runtime of around 1 hour and 15 minutes.}
    \label{fig:yHt}
\end{figure}

One may also consider approaches other than increasing the bin width and we now discuss an alternative based on the following observations. The phenomenon of misbinning introduces unphysical outliers into the set of bin value estimates. Clearly a single extreme entry can change the arithmetic mean by a large amount. Then it makes sense to employ an estimator that is robust to such extreme data. This is the topic of {\em robust statistics}, see e.g., the textbook~\cite{Maronna:2018} and references therein. Without entering into the details, let us note that in practice one of the simplest methods for building such robust estimators is via trimming. For example, the $\alpha$-{\em trimmed mean} of a sample of $n$ numbers $\{x_1, x_2, \ldots, x_n\}$ is simply the arithmetic mean of the $(n-2m)$ numbers obtained by removing the smallest and largest $m=[n \alpha]$ entries (here 
$[\ldots]$ denotes the integer part of a number),\footnote{Note that an alternate definition is sometimes used in the literature~\cite{Andrews:1972},
\[
T(x) = \frac{p x_{([\alpha n+1])} + x_{([\alpha n+2])} + \ldots + p x_{(n-[\alpha n])}}{n(1-2\alpha)}\,,
\]
where $p=1+[\alpha n]-\alpha n$. The two definitions differ only in their treatment of the largest and smallest remaining points after having removed a total of $2[\alpha n]$ points.}
\beq
\bar{x}_{\alpha} = \frac{1}{n-2m} \sum_{i=m+1}^{n-m} x_{(i)}\,,
\label{eq:al-trimmed-mean}
\eeq
where $x_{(i)}$ denotes the {\em order statistics} $\{x_{(1)}\,, x_{(2)}\,, \ldots\,, x_{(n)}\}$, obtained by sorting the $n$ numbers such that
\beq
x_{(1)} \le x_{(2)}\le \ldots \le x_{(n)}\,.
\eeq
While it may seem that such a procedure suppresses observations, in reality no subjective choice is being made: the result is actually a function of all
observations (even of those that are not included in computing the trimmed average)~\cite{Maronna:2018}. Such a trimming procedure could in principle be implemented already at the level of histogram collection in the code. However, we opt for a more straightforward solution and exploit the fact that it is very typical that final results are obtained by combining several statistically independent runs by computing bin-by-bin averages of the bin value estimates coming from the individual runs.\footnote{The data shown in \figs{fig:yH}{fig:yHt} was obtained by using 256 separate runs. We stress that the generation of the integration grids and this full set of 256 runs together requires a runtime of around 1 hour and 15 minutes on the described architecture.}
We compute bin-by-bin the $\alpha$-trimmed means with $\alpha=0.015$ (corresponding to dropping the lowest and highest 3 results for each bin) and in \fig{fig:yHt} we show the result of such a procedure. The ratios of bin values obtained from the arithmetic mean and the trimmed mean are shown as black data points in the lower panels of these plots. Let us make several observations regarding these results. First, we see that trimming is very effective in mitigating misbinning, as evidenced by comparing the plots in \figs{fig:yH}{fig:yHt}. Second, if the result was already free of misbinning, as in the case of the distributions with $\Delta y=0.25$ on the left, the trimming procedure produces basically no changes. Third, we see that the trimmed bin value estimates are compatible with the untrimmed bin values within uncertainty. Fourth, the uncertainties of the trimmed values are generally quite close to the original uncertainty estimates in regions where no misbinning is apparent, while in bins with obvious misbinning, the error estimate is strongly reduced. Last, trimming does not introduce any obvious bias and the ratio of the trimmed bin values to the untrimmed ones scatter around one with about the same number of points above and below.

\begin{figure}[t]
    \centering
 \includegraphics[width=0.4\linewidth]{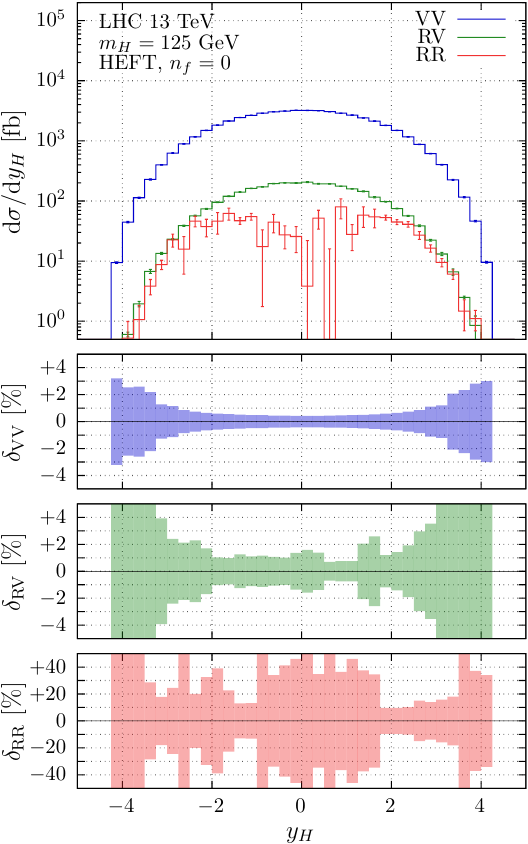}  
 \hspace{1.5em}
 \includegraphics[width=0.4\linewidth]{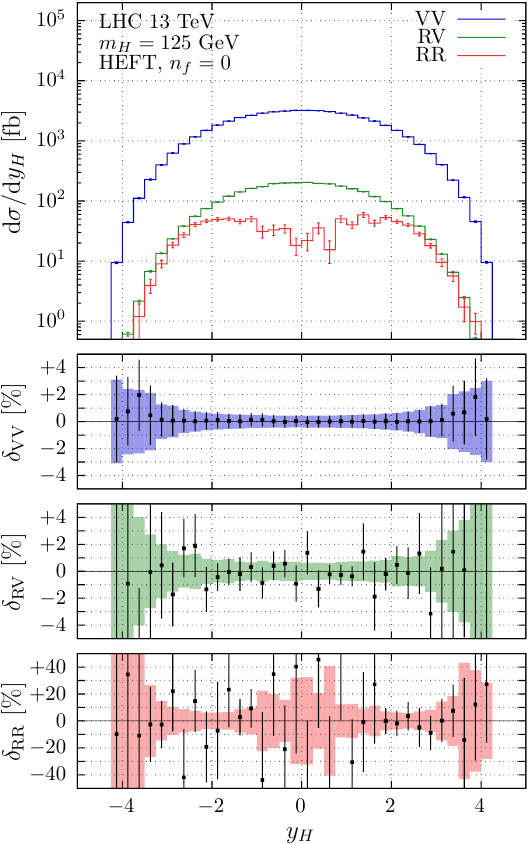}
    \caption{The double virtual (VV), real-virtual (RV) and double real (RR) parts of the NNLO correction to the Higgs boson rapidity distribution. Left: untrimmed distributions. Right: trimmed distributions. Note that the real-virtual distribution is negative and so is plotted with the opposite sign. The various plot elements are the same as in \fig{fig:yH}.}
    \label{fig:yHnnlo}
\end{figure}

Finally, we present the breakdown of the NNLO correction to the rapidity distribution into its double virtual, real-virtual and double real parts in \fig{fig:yHnnlo}. The left and right plots show the untrimmed and trimmed results. As before, the top panels show the various distributions, while the relative errors are indicated in the bottom panels. On the right plot, black data points in the bottom panels represent the ratios of untrimmed to trimmed bin values. The distributions for the double virtual and real-virtual contributions are quite stable and demonstrate good numerical convergence, while the one for the double real contribution clearly shows the impact of misbinning. However, the trimming procedure we described above mitigates this to a large extent and the relative error of the trimmed result is mostly below 20\% in the central rapidity region of $|y_H| < 2.5$. Note though that the double real piece contributes to the total NNLO result at the level of a few percent only, so this relative uncertainty translates to a permille error on the physical distribution.

%%%
%%% Conclusions and outlook
%%%

\section{Conclusions and outlook}
\label{sec:conc}

Computing higher-order perturbative corrections to collider observables is of paramount importance to fully exploit the physics potential of the LHC and future colliders. When evaluating such corrections, we must deal with various singularities that arise at intermediate stages of the computation. In particular, the straightforward application of perturbation theory is hampered by the presence of IR singularities which must be regularized and properly treated before any numerical calculation can take place. One way of dealing with IR divergences is through the use of a local subtraction scheme. Although the underlying principles of constructing such a scheme are well-understood, developing actual realizations beyond NLO have proven to be surprisingly difficult. 

In this paper, we presented the extension of the CoLoRFulNNLO method to hadron-initiated processes, concentrating on color-singlet production. Subtraction terms in this scheme are constructed directly from QCD IR limit formul\ae\ and are completely local in phase space. The corresponding integrated subtraction terms were computed fully analytically and the cancellation of all virtual poles was demonstrated explicitly. Although we focused here on color-singlet production, the obtained subtraction terms can be used to regularize initial-state radiation at NNLO also in more complicated processes. We have moreover presented a proof-of-concept implementation of our method in the code {\tt NNLOCAL}. Our code provides the first public implementation of a completely local subtraction scheme at NNLO and employs closed analytic expressions for all integrated counterterms. Concentrating on Higgs boson production at the LHC in HEFT without light quarks, we have demonstrated the viability of our approach and validated our code by a tuned comparison to publicly available tools.

This work opens the door to the application of the CoLoRFulNNLO subtraction scheme to hadron-hadron collisions. The presented code, although limited in its scope for the moment, will serve as a basis for refinements and extensions, making it a useful tool for computing NNLO QCD corrections to a plethora of important LHC processes. As a first step, we will extend our code to incorporate all partonic channels that can arise in color-singlet production, such that any color-singlet production process may be implemented by simply specifying the correct matrix elements. In this way we will be able to build a library for computing color-singlet production processes at NNLO accuracy based on a fully local and analytic subtraction scheme. The extension of our subtraction scheme to general hadronic processes is also feasible and will be the subject of further study.

%%%
%%% Acknowledgments
%%%

\section*{Acknowledgments}

This work has been supported by grant K143451 of the National Research, Development and Innovation Fund in Hungary and the Bolyai Fellowship program of the Hungarian Academy of Sciences. The work of C.D.\ was funded by the European Union (ERC Consolidator Grant LoCoMotive 101043686). Views and opinions expressed are however those of the author(s) only and do not necessarily reflect those of the European Union or the European Research Council. Neither the European Union nor the granting authority can be held responsible for them. The work of L.F.\ was supported by the German Academic Exchange Service (DAAD) through its Bi-Nationally Supervised Scholarship program. The work of F.G. is supported in part by the Swiss National Science Foundation (SNSF) under contract $200020\_219367$.

%%%
%%% Appendix
%%%

\appendix

%%%
%%% Matrix elements and insertion operators
%%%

\section{Matrix elements and insertion operators}
\label{appx:poles}

In this appendix we present the necessary ingredients to check the cancellation of $\eps$-poles in \eqn{eq:IntVVNNLO}. As stated above, we are working in the HEFT approximation without light quarks ($\Nf = 0$). Throughout this appendix, we set $\as = \as(\mu_R^2)$.

% Matrix elements

\subsection{Matrix elements}
\label{appx:mes}

The Born squared matrix element for $gg\to H$ production (in HEFT with $\Nf=0$) averaged over initial spins and colors reads
\beq
\SME{gg\to H}{(0)}{} = \frac{\as^2 m_H^4}{72\pi^2 v^2 (N_c^2-1)(1-\eps)}\,,
\eeq%
where the Higgs VEV can be expressed in terms of the Fermi constant as $v^2 = \frac{1}{\sqrt{2}G_F}$. The one-loop correction can then be written as
\beq
\bsp
\SME{gg\to H}{}{}_{1\mathrm{-loop}} &= 
2\Re \braket{gg\to H}{(0)}{}{gg\to H}{(1)}{} 
\\ &
= 
\frac{\as}{2\pi}\frac{e^{\eps \gamma_E}}{\Gamma(1-\eps)} \CA 
    \bigg[
    -\frac{2}{\ep^2} 
    - \bigg(\frac{11}{3} + 2 L_R\bigg)\frac{1}{\ep}
    + \frac{11}{3} + \pi^2 - L_R^2
    - \bigg(2 + \frac{11 \pi^2}{36} - 4 \zeta_3 
    - \pi^2 L_R     
\\ & \qquad\qquad\qquad
    + \frac{1}{3} L_R^3\bigg) \ep 
    - \bigg(6 - \frac{11 \pi^2}{36} + \frac{11 \zeta_3}{9} 
    + \frac{\pi^4}{60} + (2 - 4 \zeta_3) L_R 
    - \frac{1}{2} \pi^2 L_R^2 + \frac{1}{12} L_R^4\bigg) \ep^2 
\\ & \qquad\qquad\qquad
    + \Oe{3}
    \bigg] \SME{gg\to H}{(0)}{}\,,
\esp
\eeq
where $L_R = \ln\frac{\mu_R^2}{m_H^2}$. Finally, the two-loop correction reads
\beq
\bsp
\SME{gg\to H}{}{}_{2\mathrm{-loop}} &= 2\Re \braket{gg\to H}{(0)}{}{gg\to H}{(2)}{} + \SME{gg\to H}{(1)}{} 
\\&
= 
\left(\frac{\as}{2\pi}\right)^2\frac{e^{2\eps \gamma_E}}{\Gamma^2(1-\eps)} \CA^2 
    \bigg[
    \frac{2}{\ep^4} 
    + \bigg(\frac{121}{12} + 4 L_R\bigg)\frac{1}{\ep^3} 
    + \bigg(\frac{8}{9} - \frac{23 \pi^2}{12} 
    + \frac{55}{6} L_R + 4 L_R^2\bigg)\frac{1}{\ep^2}
\\&\qquad\qquad\qquad
    - \bigg(\frac{428}{27} + \frac{22 \pi^2}{9} 
    + \frac{15 \zeta_3}{2} 
    + \bigg(\frac{199}{18} + \frac{23 \pi^2}{6}\bigg) L_R 
    - \frac{11}{3} L_R^2 - \frac{8}{3} L_R^3\bigg)\frac{1}{\ep} 
\\&\qquad\qquad\qquad
    + \frac{15235}{324} + \frac{1961 \pi^2}{216} 
    - \frac{55 \zeta_3}{3} 
    + \frac{137 \pi^4}{360} 
    + \frac{19}{18} L_t - \bigg(\frac{176}{27} 
    - \frac{11 \pi^2}{36} 
    + 15 \zeta_3\bigg) L_R 
\\&\qquad\qquad\qquad
    - \bigg(\frac{133}{18} + \frac{23 \pi^2}{6}\bigg) L_R^2 
    + \frac{11}{18} L_R^3 + \frac{4}{3} L_R^4
    + \Oe{1}
    \bigg] \SME{gg\to H}{(0)}{}\,,
\esp
\eeq
with $L_t = \ln\frac{\mu_R^2}{m_t^2}$.

% Pole parts of insertion operators

\subsection{Pole parts of insertion operators}
\label{appx:Iops}

The expressions for the insertion operators involve Dirac-delta and plus-distributions that a priori act on differential cross sections. Reinstating the momentum dependence of the partonic cross sections for clarity, we have
\beq
\bsp
&\int_0^1 \rd x_a\, \rd x_b\, f_{a/A}(x_a) f_{b/B}(x_b) \Big(\bI(\eps) \otimes \dsig{}(x_a p_A,x_b p_B)\Big)_{ab} 
\\& =
\int_0^1 \rd x_a\, \rd x_b\, f_{a/A}(x_a) f_{b/B}(x_b) \left[\int_0^1 \rd \eta_a\, \rd \eta_b\, \bI_{ac,bd}(\eta_a,\eta_b;\eps)\, \dsig{}_{cd}(\eta_a x_a p_A,\eta_b x_b p_B)\right]\,.
\esp
\label{eq:Iopaction}
\eeq
Hence, a direct implementation of \eqn{eq:Iopaction} would require the computation of $ \dsig{}_{cd}(\eta_a x_a p_A,\eta_b x_b p_B)$ in several {\em different} phase space points. However, in a numeric calculation it is more convenient to evaluate the differential cross section in a {\em single} phase space 
point only. For this reason, we perform a change of variables from $x_a$ and $x_b$ to $\xi_a = \eta_a x_a$ and $\xi_b = \eta_b x_b$. This way, the cross section only depends on $\xi_a$ and $\xi_b$ and the action of the distributions is transferred to the product of PDFs. We then find
\beq
\bsp
&\int_0^1 \rd x_a\, \rd x_b\, f_{a/A}(x_a) f_{b/B}(x_b) \left[\int_0^1 \rd \eta_a\, \rd \eta_b\, \bI_{ac,bd}(\eta_a,\eta_b;\eps)\, \dsig{}_{cd}(\eta_a x_a p_A,\eta_b x_b p_B)\right]
\\ &=
\int_0^1 \rd \xi_a \rd \xi_b 
\int_0^1 \rd \eta_a\, \rd \eta_b\, 
\bigg[
I_{ac,bd}(\eta_a,\eta_b;\eps\,|\,\eta_a,\eta_b) \frac{f_{a/A}(\xi_a/\eta_a)}{\eta_a} \frac{f_{b/B}(\xi_b/\eta_b)}{\eta_b}
\\&\qquad
+ I_{ac,bd}(\eta_a,\eta_b;\eps\,|\,1,\eta_b) f_{a/A}(\xi_a) \frac{f_{b/B}(\xi_b/\eta_b)}{\eta_b}
+ I_{ac,bd}(\eta_a,\eta_b;\eps\,|\,\eta_a,1) \frac{f_{a/A}(\xi_a/\eta_a)}{\eta_a} f_{b/B}(\xi_b)
\\&\qquad
+ I_{ac,bd}(\eta_a,\eta_b;\eps\,|\,1,1) f_{a/A}(\xi_a) f_{b/B}(\xi_b)
\bigg]
\dsig{}_{cd}(\xi_a p_A,\xi_b p_B)\,.
\esp
\label{eq:distexplicit}
\eeq
We refer to  $I_{ac,bd}(\eta_a,\eta_b;\eps\,|\,\kappa_a,\kappa_b)$, with $\kappa_a=1,\eta_a$ and $\kappa_b=1,\eta_b$, as the coefficient functions of the operator $\bI_{ac,bd}(\eta_a,\eta_b;\eps)$. Indeed, these objects are now just functions (as opposed to distributions) of the variables $\eta_a$ and $\eta_b$. The arguments after the separator serve to specify the precise combination of PDFs that each function multiplies.\footnote{To see the correspondence with the usual notation, consider the distribution $D(x) = A\delta(1-x) + \left[\frac{B}{1-x}\right]_+ + C_{\mathrm{reg}}(x)$, where $C_{\mathrm{reg}}(x)$ is a regular function at $x=1$. Then, the action of $D(x)$ on some test function $f(x)$ can be written as
\beq
D\otimes f = 
\int_0^1 \rd x\,
\left(A\delta(1-x) + \left[\frac{B}{1-x}\right]_+ + C_{\mathrm{reg}}(x)\right)f(x) =\nn
\int_0^1 \left[\left(\frac{B}{1-x}+C_{\mathrm{reg}}(x)\right) f(x) + \left(A-\frac{B}{1-x}\right) f(1)\right]\,. 
\eeq
Thus, the integrand is a linear combination of $f(x)$ and $f(1)$ with coefficients that are simply functions of $x$. Specifying these coefficient functions, $D(x\,|\,x) = \left(\frac{B}{1-x}+C_{\mathrm{reg}}(x)\right)$ and $D(x\,|\,1) = \left(A-\frac{B}{1-x}\right)$ in this example, is an equally valid way of describing $D(x)$. This way of specifying the distribution also lends itself directly to implementation in a numerical calculation.} Below, we present the pole parts of these functions for all operators that appear in \eqn{eq:IntVVNNLO}, for the purely gluonic subprocess.

Starting with $\bI^{(0)}_{1}(\ep)$ which acts on the virtual cross section $\dsig{V}$ in \eqn{eq:IntVVNNLO}, we have
\beq
\bI^{(0)}_{1}(\ep) = \frac{\as}{2\pi}\frac{e^{\eps \gamma_E}}{\Gamma(1-\eps)} \CA \bom{\bar{I}}^{\,(0)}_{1}(\ep)\,.
\eeq
Then, the various coefficient functions of the operator $\bom{\bar{I}}^{\,(0)}_{1}(\ep)$ read
\beq
\bsp
% I10ab 
%%STARTT
\bar{I}^{\,(0)}_{1;gg,gg}(\eta_a,\eta_b;\eps\,|\,\eta_a,\eta_b) = \,&\, 2\Bigg\{\frac{-2+\eta_a-\eta_a^2}{1-\eta_b}+\frac{2}{ \eta_a \eta_b}-\frac{2+\eta_a+\eta_a^2}{1+\eta_b}-\frac{2-\eta_b+\eta_b^2}{1-\eta_a}-\frac{2+\eta_b+\eta_b^2}{1+\eta_a}\\&-\frac{1}{\eta_b\left(1+\eta_a\right)}-\frac{1}{\eta_a\left(1+\eta_b\right)}+\frac{1}{\left(1-\eta_a\right)\left(1-\eta_b\right)}+\frac{1}{\eta_a\left(1-\eta_b\right)}+\frac{1}{\eta_b\left(1-\eta_a\right)}\\&+\frac{1}{\left(1+\eta_a\right)\left(1+\eta_b\right)}+4-2 \eta_a \eta_b+2\eta_b^2+2\eta_a^2\left(1+\eta_b^2\right)\Bigg\}\\&\times\Bigg\{1+\eps\Big(2 \ln\left(\eta_a+\eta_b\right) - \ln\left(1-\eta_a\right)- \ln\left(1+\eta_a\right)- \ln\left(1-\eta_b\right)- \ln\left(1+\eta_b\right)\\&+L_R\Big)\Bigg\} + \Oe{2}\,,
%%STOPP
\esp
\label{eq:I10ab}
\eeq
\beq
\bsp
%I10a1 
%%STARTT
\bar{I}^{\,(0)}_{1;gg,gg}(\eta_a,\eta_b;\eps\,|\,\eta_a,1) = \,&\, -2p_{\text{gg}}(\eta_a)\Bigg\{\frac{1}{\eps}- \ln(2) - \ln\left(1-\eta_a\right)+ \ln\left(1+\eta_a\right)+\frac{1}{1-\eta_b}+L_R\\&+\eps\Bigg(\frac{1}{2} \ln^2(2)+\frac{1}{2} \ln^2\left(1-\eta_a\right)+ \ln\left(1-\eta_a\right)\left( \ln(2)- \ln\left(1+\eta_a\right)\right)\\&- \ln(2) \ln\left(1+\eta_a\right)+\frac{1}{2} \ln^2\left(1+\eta_a\right)+\frac{1}{2}L_R ^2-L_R \left( \ln(2)+ \ln\left(1-\eta_a\right)- \ln\left(1+\eta_a\right)\right)\\&-\frac{ \ln(2) + \ln\left(1-\eta_a\right)- \ln\left(1+\eta_a\right)+ \ln\left(1-\eta_b\right)-L_R}{1-\eta_b}
\Bigg)\Bigg\} + \Oe{2}\,,
%%STOPP
\esp
\label{eq:I10a1}
\eeq
\beq
\bsp
%I1011 
%%STARTT
\bar{I}^{\,(0)}_{1;gg,gg}(\eta_a,\eta_b;\eps\,|\,1,1) = \,&\, \frac{2}{\eps^2}+\frac{2}{\eps}\Bigg\{\frac{1}{1-\eta_a}+\frac{1}{1-\eta_b}+L_R\Bigg\}+2\Bigg\{\frac{L_R^2}{2}-\frac{ \ln\left(1-\eta_a\right)-L_R}{1-\eta_a}\\&-\frac{ \ln\left(1-\eta_b\right)-L_R }{1-\eta_b}+\frac{1}{\left(1-\eta_a\right)\left(1-\eta_b\right)}\Bigg\}+\eps\Bigg\{\frac{L_R ^3}{3}+\frac{\left( \ln\left(1-\eta_a\right)-L_R\right)^2}{1-\eta_a}\\&+\frac{\left( \ln\left(1-\eta_b\right)-L_R\right)^2}{1-\eta_b}-\frac{2\left( \ln\left(1-\eta_a\right)+ \ln\left(1-\eta_b\right)-L_R\right)}{\left(1-\eta_a\right)\left(1-\eta_b\right)}
\Bigg\} + \Oe{2}
%%STOPP
\esp
\label{eq:I1011}
\eeq
where 
\beq
p_{gg}(\eta) = \frac{1}{1 - \eta} + \frac{1}{\eta} - 2 + \eta (1 - \eta)
\eeq
and $L_F = \ln\frac{\mu_F^2}{m_H^2}$. Moreover $\bar{I}^{\,(0)}_{1;gg,gg}(\eta_a,\eta_b;\eps\,|\,1,\eta_b)$ is obtained by exchanging $\eta_a$ and $\eta_b$ in the expression in \eqn{eq:I10a1},
\beq
\bar{I}^{\,(0)}_{1;gg,gg}(\eta_a,\eta_b;\eps\,|\,1,\eta_b) = 
\bar{I}^{\,(0)}_{1;gg,gg}(\eta_b,\eta_a;\eps\,|\,\eta_a,1)\,.
\eeq
For the fully gluonic subprocess, this follows from the obvious symmetry between incoming partons. Similar relations will thus hold for all other operators as well.

Next, consider the sum of operators acting on the Born cross section $\dsig{B}$ in \eqn{eq:IntVVNNLO}. It turns out that the structure of this sum is simpler than the structure of the individual operators, so we introduce
\beq
\bI_{\mathrm{B}}(\ep) = \bI^{(0)}_{2}(\ep) - \bI^{(0)}_{12}(\ep) + \bI^{(1)}_{1}(\ep) + \bI^{(0,0)}_{\Gamma,1}(\ep) + \bI^{(0,0)}_{1,1}(\ep)\,.
\eeq
This sum of operators can be written as
\beq
\bI_{\mathrm{B}}(\ep) = \left(\frac{\as}{2\pi}\right)^2\frac{e^{2\eps \gamma_E}}{\Gamma^2(1-\eps)} \CA^2 \bom{\bar{I}}_{\mathrm{B}}(\ep)
\eeq
and the coefficient functions of $\bom{\bar{I}}_{\mathrm{B}}(\ep)$ are given by
\beq
\bsp
%Itotab 
%%STARTT
\bar{I}_{\mathrm{B};gg,gg}(\eta_a,\eta_b;\eps\,|\,\eta_a,\eta_b) = \,&\, \frac{4}{\eps^2}\Bigg\{\frac{2-\eta_a+\eta_a^2}{\eta_b}-\frac{2+\eta_a+\eta_a^2}{1+\eta_b}+\frac{2-\eta_b+\eta_b^2}{\eta_a}-\frac{2+\eta_b+\eta_b^2}{1+\eta_a}\\&-\frac{1}{\eta_b\left(1+\eta_a\right)}-\frac{1}{\eta_a\left(1+\eta_b\right)}+\frac{1}{ \eta_a \eta_b}+\frac{1}{\left(1+\eta_a\right)\left(1+\eta_b\right)}\\&+\left(2\eta_b+\eta_a^2\eta_b\left(1+\eta_b\right)+\eta_a\left(2-3\eta_b+\eta_b^2\right)\right)\Bigg\}\\&
    +\frac{4}{\eps}\Bigg\{\Bigg(\frac{11}{6}- \ln\left(1-\eta_a\right)- \ln\left(1+\eta_a\right)- \ln\left(1-\eta_b\right)- \ln\left(1+\eta_b\right)\\&+2 \ln\left(\eta_a+\eta_b\right)+2L_R\Bigg)\Bigg(\frac{-2+\eta_a-\eta_a^2}{1-\eta_b}+\frac{2}{ \eta_a \eta_b}-\frac{2+\eta_a+\eta_a^2}{1+\eta_b}\\&-\frac{2-\eta_b+\eta_b^2}{1-\eta_a}-\frac{2+\eta_b+\eta_b^2}{1+\eta_a}+\frac{1}{\left(1-\eta_a\right)\left(1-\eta_b\right)}+\frac{1}{\eta_a\left(1-\eta_b\right)}\\&+\frac{1}{\eta_b\left(1-\eta_a\right)}-\frac{1}{\eta_b\left(1+\eta_a\right)}-\frac{1}{\eta_a\left(1+\eta_b\right)}+\frac{1}{\left(1+\eta_a\right)\left(1+\eta_b\right)}\\&+2\left(2- \eta_a \eta_b+\eta_b^2+\eta_a^2\left(1+\eta_b^2\right)\right)\Bigg)-2 p_{\text{gg}}(\eta_a)p_{\text{gg}}(\eta_b)L_F
\Bigg\} + \Oe{0}\,,
%%STOPP
\esp
\label{eq:Itotab}
\eeq
\beq
\bsp
%Itota1 
%%STARTT
\bar{I}_{\mathrm{B};gg,gg}(\eta_a,\eta_b;\eps\,|\,\eta_a,1) = \,&\, \frac{1}{2\eps^2}\Bigg\{\frac{-11+4 \ln\left(\eta_a\right)}{1-\eta_a}+\frac{11+12 \ln\left(\eta_a\right)}{3\eta_a}+\frac{1}{3}\Big(30+3\eta_a+36 \eta_a \ln\left(\eta_a\right)-11\eta_a^2\\&-12 \eta_a^2 \ln\left(\eta_a\right)\Big)+8\left( \ln\left(2\right)- \ln\left(1+\eta_a\right)+L_F-L_R\right)p_{\text{gg}}(\eta_a)\Bigg\}\\&
    +\frac{1}{3\eps}\Bigg\{\frac{-67+9\pi^2-9 \ln^2\left(\eta_a\right)-132L_F +72  \ln\left(\eta_a\right)L_F}{6\left(1-\eta_a\right)}\\&+\frac{\pi^2+22L_F +12  \ln\left(\eta_a\right)L_F}{\eta_a}+\frac{\pi^2-3 \ln^2\left(\eta_a\right)}{2\left(1+\eta_a\right)}+\frac{1}{12}\Big(25-48\pi^2 +150 \ln\left(\eta_a\right)\\&+109\eta_a+12\pi^2\eta_a-66 \eta_a \ln\left(\eta_a\right)-72 \eta_a \ln^2\left(\eta_a\right)-24\pi^2\eta_a^2+264 \eta_a^2 \ln\left(\eta_a\right)\\&+36 \eta_a^2 \ln^2\left(\eta_a\right)+96L_F+168 \eta_a L_F-264 \eta_a^2 L_F+432  \eta_a \ln\left(\eta_a\right) L_F\\&-144  \eta_a^2 \ln\left(\eta_a\right) L_F\Big)-2(11-12L_F )\Bigg(\frac{1}{\left(1-\eta_a\right)\left(1-\eta_b\right)}+\frac{1}{\eta_a\left(1-\eta_b\right)}\\&-\frac{2-\eta_a+\eta_a^2}{1-\eta_b}\Bigg)+6\left( \ln\left(\eta_a\right) \ln\left(1+\eta_a\right)+\Li_2\left(-\eta_a\right)\right)p_{\text{gg}}(-\eta_a)\\&-\Bigg(2\Big(3 \ln^2(2)-11 \ln(2)-11 \ln\left(1-\eta_a\right)+6 \ln(2) \ln\left(1-\eta_a\right)\\&+3 \ln^2\left(1-\eta_a\right)-3 \ln\left(1-\eta_a\right) \ln\left(\eta_a\right)+11 \ln\left(1+\eta_a\right)-6 \ln(2) \ln\left(1+\eta_a\right)\\&-6 \ln\left(1-\eta_a\right) \ln\left(1+\eta_a\right)+3 \ln^2\left(1+\eta_a\right)-3L_F ^2-6 L_R L_F+12  \ln\left(1-\eta_a\right)L_F\\&+9L_R ^2+11L_R -12 \ln(2)L_R  -12  \ln\left(1-\eta_a\right)L_R+12  \ln\left(1+\eta_a\right)L_R\Big)\\&+\frac{12\left( \ln\left(1+\eta_a\right)- \ln(2) - \ln\left(1-\eta_a\right)- \ln\left(1-\eta_b\right)+2L_R\right)}{1-\eta_b}\Bigg)p_{\text{gg}}(\eta_a)
\Bigg\} + \Oe{0}\,,
%%STOPP
\esp
\label{eq:Itota1}
\eeq
\beq
\bsp
%Itot11 
%%STARTT
\bar{I}_{\mathrm{B};gg,gg}(\eta_a,\eta_b;\eps\,|\,1,1) = \,&\, \frac{2}{\eps^4}+\frac{1}{\eps^3}\Bigg\{\frac{55}{12}+4 L_R\Bigg\}+\frac{1}{2\eps^2}\Bigg\{\frac{1}{18}\left(67+21\pi^2+264L_F+144L_R ^2 +198L_R \right)\\&+(11-8L_F +8L_R )\left(\frac{1}{1-\eta_a}+\frac{1}{1-\eta_b}\right)\Bigg\}\\&+\frac{1}{18\eps}\Bigg\{\frac{202}{3}-63\zeta_3+66L_F ^2+24\pi^2L_F+48L_R ^3+66L_R ^2 +67L_R -3\pi^2L_R \\&+132L_F L_R -\frac{1}{1-\eta_a}\Big(-67+9\pi^2+36L_F ^2-132L_F -144  \ln\left(1-\eta_a\right)L_F \\&+72 L_R L_F -108L_R ^2-132L_R+144  \ln\left(1-\eta_a\right)L_R+132 \ln\left(1-\eta_a\right)\\&-36 \ln^2\left(1-\eta_a\right)\Big)-\frac{1}{1-\eta_b}\Big(-67+9\pi^2+36L_F ^2-132L_F\\&-144  \ln\left(1-\eta_b\right)L_F +72 L_R L_F -108L_R ^2-132L_R+144  \ln\left(1-\eta_b\right)L_R \\&+132 \ln\left(1-\eta_b\right)-36 \ln^2\left(1-\eta_b\right)\Big)+\frac{72}{(1-\eta_a)(1-\eta_b)}\Bigg(\frac{11}{6}- \ln\left(1-\eta_a\right)\\&- \ln\left(1-\eta_b\right)-2L_F +2L_R \Bigg)
\Bigg\} + \Oe{0}\,.
%%STOPP
\esp
\label{eq:Itot11}
\eeq
As explained above, here too, we have
\beq
\bar{I}_{\mathrm{B};gg,gg}(\eta_a,\eta_b;\eps\,|\,1,\eta_b) = 
\bar{I}_{\mathrm{B};gg,gg}(\eta_b,\eta_a;\eps\,|\,\eta_a,1)\,.
\eeq

Finally, although the expressions of $\bom{\Gamma}$ are well-known, for completeness we present these explicitly as well. Again, it is useful to extract powers of $\as$ and $\CA$, and we set
\beq
\bom{\Gamma}^{(1)} = \frac{\as}{2\pi} \CA \bom{\overline{\Gamma}}^{(1)}
\qquad\mbox{and}\qquad
\bom{\Gamma}^{(2)} = \left(\frac{\as}{2\pi}\right)^2 \CA^2 \bom{\overline{\Gamma}}^{(2)}\,.
\eeq
The corresponding coefficient functions then read
\beq
\bsp
%Gam1ab 
%%STARTT
\overline{\Gamma}^{(1)}_{gg,gg}(\eta_a,\eta_b;\eps\,|\,\eta_a,\eta_b) = \,&\, 0\,,
%%STOPP
\esp
\label{eq:G1ab}
\eeq
\beq
\bsp
%Gam1a1 
%%STARTT
\overline{\Gamma}^{(1)}_{gg,gg}(\eta_a,\eta_b;\eps\,|\,\eta_a,1) = \,&\, 2 p_{\text{gg}}(\eta_a)\Bigg\{\frac{1}{\eps}+L_F+\eps\frac{L_F^2}{2}\Bigg\} + \Oe{2}\,,
%%STOPP
\esp
\label{eq:G1a1}
\eeq
\beq
\bsp
%Gam111 
%%STARTT
\overline{\Gamma}^{(1)}_{gg,gg}(\eta_a,\eta_b;\eps\,|\,1,1) = \,&\, \Bigg\{\frac{11}{3}-\frac{2}{1-\eta_a}-\frac{2}{1-\eta_b}\Bigg\}\Bigg\{\frac{1}{\eps}+L_F+\eps\frac{L_F^2}{2}\Bigg\} + \Oe{2}
%%STOPP
\esp
\label{eq:G111}
\eeq
and 
\beq
\bsp
%Gam2ab 
%%STARTT
\overline{\Gamma}^{(2)}_{gg,gg}(\eta_a,\eta_b;\eps\,|\,\eta_a,\eta_b) = \,&\, 4p_{\text{gg}}(\eta_a)p_{\text{gg}}(\eta_b)\Bigg\{\frac{1}{\eps^2}+\frac{2L_F}{\eps}\Bigg\} + \Oe{0}\,,
%%STOPP
\esp
\label{eq:G2ab}
\eeq
\beq
\bsp
%Gam2a1 
%%STARTT
\overline{\Gamma}^{(2)}_{gg,gg}(\eta_a,\eta_b;\eps\,|\,\eta_a,1) = \,&\, \frac{1}{\eps^2}\Bigg\{\frac{11-4 \ln\left(\eta_a\right)}{2\left(1-\eta_a\right)}-\frac{11+12 \ln\left(\eta_a\right)}{6\eta_a}-5-\frac{1}{2}\eta_a-6\eta_a \ln\left(\eta_a\right)+\frac{11}{6}\eta_a^2\\&+2\eta_a^2 \ln\left(\eta_a\right)-\frac{4}{\left(1-\eta_a\right)\left(1-\eta_b\right)}-\frac{4}{\eta_a\left(1-\eta_b\right)}+\frac{4\left(2-\eta_a+\eta_a^2\right)}{1-\eta_b}\\&+4 \ln\left(1-\eta_a\right)p_{\text{gg}}(\eta_a)\Bigg\}\\&
    +\frac{1}{\eps}\Bigg\{\frac{67-3\pi^2+9 \ln^2\left(\eta_a\right)+264L_F -72  \ln\left(\eta_a\right)L_F}{18\left(1-\eta_a\right)}-\frac{\pi^2-3 \ln^2\left(\eta_a\right)}{6\left(1+\eta_a\right)}\\&+\frac{1}{36}
    (-25+24\pi^2 -150 \ln\left(\eta_a\right)-109\eta_a+66\eta_a \ln\left(\eta_a\right)+72\eta_a \ln^2\left(\eta_a\right)\\&+12\pi^2\eta_a^2-264\eta_a^2 \ln\left(\eta_a\right)-36\eta_a^2 \ln^2\left(\eta_a\right)-624L_F+96\eta_a L_F\\&-432\eta_a \ln\left(\eta_a\right)L_F+144\eta_a^2  \ln\left(\eta_a\right) L_F)-\frac{4L_F}{\eta_a(1-\eta_a)(1-\eta_b)}(2+ \ln\left(\eta_a\right)\\&-4\eta_a-\eta_a \ln\left(\eta_a\right)+6\eta_a^2-4\eta_a^3+2\eta_a^4-\eta_b \ln\left(\eta_a\right)+\eta_a \eta_b \ln\left(\eta_a\right))\\&-2\left( \ln\left(\eta_a\right) \ln\left(1+\eta_a\right)+\Li_2\left(-\eta_a\right)\right)p_{\text{gg}}(-\eta_a)\\&-2 \ln\left(1-\eta_a\right)\left(  \ln\left(\eta_a\right)-4L_F\right)p_{\text{gg}}(\eta_a)
\Bigg\} + \Oe{0}\,,
%%STOPP
\esp
\label{eq:G2a1}
\eeq
\beq
\bsp
%Gam211 
%%STARTT
\overline{\Gamma}^{(2)}_{gg,gg}(\eta_a,\eta_b;\eps\,|\,1,1) = \,&\, \frac{1}{2\eps^2}\Bigg\{\frac{1}{18}\left(121-24\pi^2\right)-\frac{11+8 \ln\left(1-\eta_a\right)}{1-\eta_a}-\frac{11+8 \ln\left(1-\eta_b\right)}{1-\eta_b}\\&+\frac{8}{\left(1-\eta_a\right)\left(1-\eta_b\right)}\Bigg\}\\&+\frac{1}{9\eps}\Bigg\{24+27\zeta_3+121L_F -12\pi^2L_F -\frac{1}{2(1-\eta_a)}(67-3\pi^2\\&+264L_F +144  \ln\left(1-\eta_a\right)L_F)-\frac{1}{2(1-\eta_b)}(67-3\pi^2+264L_F \\&+144  \ln\left(1-\eta_b\right)L_F)+\frac{72L_F }{\left(1-\eta_a\right)\left(1-\eta_b\right)}
\Bigg\} + \Oe{0}\,.
%%STOPP
\esp
\label{eq:G211}
\eeq
Of course, as before
\beq
\overline{\Gamma}^{(1)}_{gg,gg}(\eta_a,\eta_b;\eps\,|\,1,\eta_b) = 
\overline{\Gamma}^{(1)}_{gg,gg}(\eta_b,\eta_a;\eps\,|\,\eta_a,1)
\eeq
and
\beq
\overline{\Gamma}^{(2)}_{gg,gg}(\eta_a,\eta_b;\eps\,|\,1,\eta_b) = 
\overline{\Gamma}^{(2)}_{gg,gg}(\eta_b,\eta_a;\eps\,|\,\eta_a,1)\,.
\eeq

To finish, let us make the following point regarding our implementation of \eqn{eq:IntVVNNLO}. Since the combination of terms in this equation does not have $\eps$-poles, in a numerical computation in four dimensions the full expression may be multiplied freely with any $\eps$-dependent constant $C(\eps)$ of the form $C(\eps) = 1 + \Oe{}$. Clearly this does not influence the four-dimensional value of the expression. In our concrete implementation, we include such a factor given by 
\beq
C(\eps) = \left[\frac{e^{\eps \gamma_E}}{\Gamma(1-\eps)}\right]^{-2}
\frac{\SME{gg\to H}{(0)}{}_{\ep=0}}{\SME{gg\to H}{(0)}{}}\,.
\eeq
In practice, we achieve this overall multiplication by setting (notice $\SME{gg\to H}{(0)}{}_{\ep=0}/\SME{gg\to H}{(0)}{} = 1-\eps$)
\beq
\SME{gg\to H}{}{}_{l\mathrm{-loop}} \to (1-\eps)
\left[\frac{e^{\eps \gamma_E}}{\Gamma(1-\eps)}\right]^{-l}\SME{gg\to H}{}{}_{l\mathrm{-loop}}\,,\qquad l=0,1,2\,,
\eeq
(in particular, the Born matrix element is normalized to its four-dimensional value), and
\bal
\bI^{(0)}_{1}(\eps) &\to \left[\frac{e^{\eps \gamma_E}}{\Gamma(1-\eps)}\right]^{-1} \bI^{(0)}_{1}(\eps)
= \frac{\as}{2\pi}\CA \bom{\bar{I}}^{\,(0)}_{1}(\eps)\,,
\\
\bI_{\mathrm{B}}(\eps) &\to \left[\frac{e^{\eps \gamma_E}}{\Gamma(1-\eps)}\right]^{-2} \bI_{\mathrm{B}}(\eps)
= \left(\frac{\as}{2\pi}\right)^2 \CA^2 \bom{\bar{I}}_{\mathrm{B}}(\eps)\,,
\\
\bom{\Gamma}^{(1)} &\to \left[\frac{e^{\eps \gamma_E}}{\Gamma(1-\eps)}\right]^{-1} \bom{\Gamma}^{(1)} 
= \frac{\as}{2\pi}\CA \left[\frac{e^{\eps \gamma_E}}{\Gamma(1-\eps)}\right]^{-1} \bom{\overline{\Gamma}}^{(1)}\,,
\\
\bom{\Gamma}^{(2)} &\to \left[\frac{e^{\eps \gamma_E}}{\Gamma(1-\eps)}\right]^{-2} \bom{\Gamma}^{(2)} 
= \left(\frac{\as}{2\pi}\right)^2\CA^2 \left[\frac{e^{\eps \gamma_E}}{\Gamma(1-\eps)}\right]^{-2} \bom{\overline{\Gamma}}^{(2)}\,
\eal
in our implementation.

%%%
%%% Installing and running {\tt NNLOCAL}
%%%

\section{Installing and running {\tt NNLOCAL}}
\label{app:install}

Our code can be obtained at \url{https://github.com/nnlocal/nnlocal.git}. After cloning the git repository into the desired directory, the code can be compiled with the included {\tt makefile} by running {\tt make}. The only external dependency is LHAPDF~\cite{Buckley:2014ana}. After compilation, the executable {\tt nnlocal} is created in the {\tt bin} directory. A run can then be set up by editing the provided {\tt input.DAT} file in the {\tt bin/testrun-H} subdirectory. The most important parameters that are set in this file are the following. 
\begin{itemize}
    \item {\tt nproc}: the process ID number. Currently only the $pp \to H$ process is implemented, for which {\tt nproc~=~710}.
    \item {\tt order}: the order in $\as$ relative to the Born process. Hence, {\tt order = 0,1,2} correspond to the LO, NLO and NNLO computations. Moreover, if {\tt order} is negative, only the correction at the appropriate relative order is computed, so {\tt order = -1,-2} give the pure NLO and NNLO contributions.
    \item {\tt part}: specifies which part of the full computation to perform. Possible values are the following. 
    \begin{enumerate}
        \item {\tt born}: include all contributions up to the given {\tt order} that have Born-like (i.e., $2\to 1$) kinematics, e.g., the double virtual contribution at NNLO; 
        \item {\tt virt}: include all contributions up to the given {\tt order} that have Born + one parton kinematics, e.g., the real-virtual contribution at NNLO; 
        \item {\tt real}: include all contributions up to the given {\tt order} that have Born + two parton kinematics, e.g., the double real contribution at NNLO; 
        \item {\tt tota}: include all contributions up to the given {\tt order}.
    \end{enumerate}
    \item {\tt sqrts}: the total center of mass energy of the hadron-hadron collision in GeV.
    \item {\tt hmass}: the mass of the Higgs boson $m_H$ in GeV.
    \item {\tt scale}: the renormalization scale $\mu_R$ in GeV.
    \item {\tt facscale}: the factorization scale $\mu_F$ in GeV.
    \item {\tt itmx1}: the total number of iterations used for grid
      refinement in serial running mode.
    \item {\tt ncall1}: the total number of evaluations per iteration
      during grid refinement.
    \item {\tt itmx2}: the total number of iterations for collecting
      results after the grid has been set up.
    \item {\tt ncall2}: the total number of evaluations per iteration
      during result collection.
    \item {\tt (ncall virt)/(ncall born)}: if {\tt part = tota} is
      selected, then all partonic contributions are evaluated during
      the run. In this case, {\tt ncall1} and {\tt ncall2} give the
      number of evaluations per iteration during the grid refinement
      and collection stages for the contributions with Born-like
      kinematics. However it is usually necessary to run
      higher-multiplicity partonic processes with more points and this
      parameter provides a way to increase the total number of points
      by multiplying {\tt ncall1} and {\tt ncall2} with the value set
      here for contributions with Born + one parton kinematics. If the
      value of {\tt part} is something other than {\tt tota}, this
      parameter is inactive.
    \item {\tt (ncall real)/(ncall born)}: same as above, for Born +
      two parton kinematics.
    \item {\tt parallel}: specify whether to run in serial mode ({\tt
      0}) or parallel mode ({\tt 1}), see below.
\end{itemize}

After setting up the inputs, the code can be run in serial mode (with
the {\tt parallel} flag set to {\tt 0}) by simply invoking the
executable. Assuming the input file is prepared in a subdirectory of
{\tt bin}, we have
\begin{verbatim}
    ../nnlocal <file>
\end{verbatim}
Here the optional argument {\tt <file>} allows to use a file different
from the default {\tt input.DAT} for specifying the run
parameters. With this setup, the integration grids are first refined
over ${\tt itmx1}$ iterations, then in a second stage, results are
gathered with fixed integration grids. In this second stage, a
user-defined analysis routine is also invoked for each event allowing
e.g., the collection of histograms for physical observables.

Our code can also be run in parallel mode with the help of the
included scripts. The parallelization is achieved in a very
straightforward manner and is built to exploit architectures with
several CPU cores. To perform a parallel run, one must first set the
{\tt parallel} flag to {\tt 1} in the input card. The launching of
jobs and collection of results is then controlled through the {\tt
  runpar.sh} script. The most important variables in this script are
\begin{itemize}
    \item {\tt ncores}: the number of CPU cores that the user wishes to use simultaneously.
    \item {\tt nprocessesgrid}: the total number of instances to be used during each iteration step of grid refinement.
    \item {\tt nprocessesaccu}: the total number of instances to be used during the collection of results after the grid has been set up.
    \item {\tt maxgrid}: the number of iteration steps used for grid refinement.
    \item {\tt alpha}: the trimming parameter $\alpha$ introduced in \sect{sec:nnlocal}.
\end{itemize}
After setting these parameters, the script will launch batches of
${\tt ncores}$ jobs as necessary to produce the total number of runs
specified. The execution in the parallel setup also proceeds in two
stages. In the first stage, {\tt nprocessesgrid} jobs are completed in
batches of ${\tt ncores}$.\footnote{The total number of evaluations in
each job is still set by {\tt ncall1}, however the value set for {\tt
  itmx1} is now irrelevant, as the number of iterations for grid
refinement is controlled by {\tt maxgrid}.} After all jobs are
finished, the obtained separate integration grids are averaged to
produce a single grid for the next batch of runs. In total, {\tt
  maxgrid} steps of grid refinement are performed in this manner,
which completes the first stage of running. Then, during the second
stage, a total of {\tt nprocessesaccu} jobs are launched in batches of
{\tt ncores}.\footnote{The total number of evaluations for each job
during this stage is set by {\tt ncall2} {\em and} {\tt itmx2}: each
run will perform {\tt itmx2} iterations of {\tt ncall2} evaluations.}
When all jobs are complete, the separate results are collected into
the output files {\tt nnlocal-1.top}, {\tt nnlocal-2.top} and {\tt nnlocal-\$m-\$m.top}, where 
{\tt \$m} is value of the integer $m$ introduced in \eqn{eq:al-trimmed-mean}.
The first file contains results obtained by computing an arithmetic average of
the separate runs, the second one contains results obtained by
computing weighted averages, while the last file contains results computed with 
the $\alpha$-trimmed mean. If the statistics are high enough, the results computed 
with the arithmetic and weighted averages should be in good agreement, hence any 
large discrepancies can be used to diagnose a situation where the complete
statistics was insufficient to produce results that have properly converged.

%%%
%%% Bibliography
%%%
% ========== ========== ========== ========== ==========

\bibliographystyle{JHEP}
\bibliography{nnlocal}

\end{document}